\documentclass[8pt,preprint2]{aastex}
\newcommand{\hst}{{\sl HST}}
\newcommand{\omcen}{{$\omega$ Cen}}

\slugcomment{version of 16 Jan. 2007}

\shorttitle{Multiplicity of \omcen\ SGB region}
\shortauthors{Villanova et al.}

\begin{document}
    \title{The multiplicity of the subgiant branch of $\omega$
    Centauri: Evidence for prolonged star formation\thanks{Based
    on FLAMES+GIRAFFE@VLT  observations under
    the  DDT program  272.D-5065(A), and  on  observations  with the
    NASA/ESA  {\it  Hubble  Space  Telescope}, obtained  at  the  Space
    Telescope Science Institute, which is operated by AURA, Inc., under
    NASA contract NAS 5-26555} }

\author{S.\ Villanova\altaffilmark{1},
        G.\ Piotto\altaffilmark{1},
        I.\ R. King \altaffilmark{2},
        J.\ Anderson\altaffilmark{3},
        L.\ R.\ Bedin\altaffilmark{4},
        R.\ G.\ Gratton\altaffilmark{6},
        S.\ Cassisi\altaffilmark{5},
        Y.\ Momany\altaffilmark{1,6},
        A.\ Bellini\altaffilmark{1},
        A.\ M.\ Cool\altaffilmark{7},
        A.\ Recio-Blanco\altaffilmark{8}, and
        A.\ Renzini\altaffilmark{6}
        }

\altaffiltext{1}{Dipartimento di Astronomia, Universit\`a di Padova,
                 Vicolo dell'Osservatorio 3, I-35122 Padua, Italy\\
                \email{sandro.villanova@unipd.it, giampaolo.piotto@unipd.it,
                       momany@pd.astro.it, andrea.bellini@unipd.it}}
\altaffiltext{2}{Department of Astronomy, University of Washington, Box
351580,
                 Seattle, WA 98195-1580\\
                 \email{king@astro.washington.edu}}
\altaffiltext{3}{Department of Physics and Astronomy, Mail Stop 108, Rice
                 University, 6100 Main Street, Houston, TX 77005\\
                 \email{jay@eeyore.rice.edu}}
\altaffiltext{4}{European Southern Observatory, Karl-Schwarzschild-Strasse 2,
                 D-85748 Garching, Germany\\
                \email{lbedin@eso.org, arenzini@eso.org}}
\altaffiltext{5}{INAF-Osservatorio Astronomico di Collurania, via Mentore
                 Maggini, 64100 Teramo, Italy\\
                 \email{cassisi@te.astro.it}}
\altaffiltext{6}{INAF-Osservatorio Astronomico di Padova, Vicolo
                 dell'Osservatorio 5, I-35122 Padua, Italy\\
                 \email{gratton@pd.astro.it, arenzini@pd.astro.it}}
\altaffiltext{7}{Department of Physics and Astronomy, San Francisco State
                 University, 1600 Holloway Avenue, San Francisco, CA 94132\\
                 \email{cool@sfsu.edu}}
\altaffiltext{8}{Observatoire de la C\^{o}te d'Azur, Dpt.\ Cassiopée UMR 6202
                 B.P.\ 4229, 06304 Nice Cedex 4, France\\
                 \email{arecio@obs-nice.fr}}

\begin{abstract}
%%%%%%%%%%%%%%%%%%%%%%%%%%%%%%%%%%%%%%%%%%%%%%%%%%%%%%%%%%%%%%%%%%%%%%%%%%%%%%
% context heading (optional) % {} leave it empty if necessary
  {We present metallicity measurements based on GIRAFFE@VLT spectra of
   80 subgiant-branch stars of the Galactic globular cluster
   $\omega$~Centauri. The VLT spectroscopic data are complemented by
   color-magnitude diagrams from high-accuracy photometry on a $\sim
   10\times10$ arcmin$^2$ mosaic of ACS/{\sl HST} images centered on the
   cluster center, and on multicolor images of a $\sim 34\times33$
   arcmin$^2$ field, taken with the WFI@ESO2.2m camera.}
% aims heading (mandatory)
  {Our main purpose was
to combine photometric data with spectroscopic data, in the hope of
teasing apart some of the population mysteries that neither
data set can answer on its own.  }
% methods heading (mandatory)
  {We have obtained the [Fe/H] abundance for each of the 80 target
   stars, and the abundances of C, N, Ca, Ti, and Ba for a subset of
   them, by comparison with synthetic spectra. We show that stars with
   [Fe/H] $<-1.25$ have a large magnitude spread on the flat part of the
   SGB. We interpret this as empirical evidence for an age spread.  A
   relative age has been obtained for each star, from theoretical
   isochrones for its metallicity, $\alpha$-enhancement, and presumed He
   content.}
% results heading (mandatory)
  {We have identified four distinct stellar groups within the SGB
   region: (i) an old, metal-poor group ([Fe/H] $\sim-1.7$);
   (ii) an old, metal-rich group ([Fe/H] $\sim-1.1$);
   (iii) a young (up to 4--5 Gyr younger than the old component)
   metal-poor group ([Fe/H] $\sim-1.7$);
   (iv) a young, intermediate-metallicity ([Fe/H] $\sim-1.4$) group,
   on average 1--2 Gyr younger than the old
   metal-poor population, and with an age spread that we cannot
   properly quantify with the present sample.  In addition,
   a group of SGB stars are spread between the intermediate-metallicity and
   metal-rich branches of the SGB.  We tentatively propose connections
   between the SGB stars and
   both the multiple main sequence and the red giant branch.
   Finally, we
   discuss the implications of the multiple stellar populations on the
   formation and evolution of $\omega$~Cen.  The spread in age within
   each population establishes that the original system must have had a
   composite nature. }
\end{abstract}


\clearpage
%
%________________________________________________________________

%%%%%%%%%%%%%%%%%%%%%%%%%%%%%%
%
\section{Introduction}
%
%%%%%%%%%%%%%%%%%%%%%%%%%%%%%%

Omega Centauri is a peculiar and enigmatic object:\ it appears to be a
globular cluster (GC), but it has a complex stellar population, and
with its unusual mass ($M\sim3\times10^6M_{\odot}$) it has often been
suggested to be the remains of a larger stellar system.  It has
received a large amount of attention; for a review see Meylan 2003.
The most provocative recent result (Anderson 1997, Bedin et al.\ 2004,
hereafter B04) was the discovery that over a range of at least two
magnitudes the main sequence splits into a red branch and a blue one.
Follow-up spectroscopic studies at medium resolution led to even more
enigmatic results (Piotto et al.\ 2005, hereafter P05):\ contrary to
any expectation from canonical stellar models, the bluer branch of the
MS is more metal-rich than the red one.  At the moment, the only
explanation of the photometric and spectroscopic properties of the
double main sequence that is at all plausible is that the bluer branch
of the MS has an unusually high helium content (B04, Norris 2004, P05,
Lee et al.\ 2005).

It has been suggested that this unusual He-rich population might come
from material contaminated by the ejecta of massive ($25 M_{\odot}$,
Norris 2004), or slightly less massive (10--14 $M_{\odot}$, P05)
supernovae, or from rapidly rotating low-metallicity massive stars
(Maeder \& Meynet 2006), or from intermediate-mass
asymptotic-giant-branch stars (Izzard et al.\ 2004).  None of these
hypotheses has been directly supported by observation.  A detailed study
of the chemical abundances of the different populations identified in
$\omega$~Cen is badly needed in order better to understand the complex
star-formation history of this cluster.

In the present paper we continue the photometric and spectroscopic
investigation started in B04 and P05, giving further results on the
double main sequence but concentrating our main attention on the region
of the subgiant branch (SGB), which is even more complex than the MS,
because of the presence of many different SGBs (Ferraro et al.\ 2004,
B04).
It is the combination of spectroscopy of SGB stars with high-accuracy
photometry that
will allow us to shed new light on the star-formation history of
$\omega$~Cen.

In Section~2 we present the photometric data. The new color-magnitude
diagrams (CMDs) are shown and discussed in Section~3, while the
spectroscopic data are presented in Section~4.  Section~5 deals with the
terminology of populations.  In Section~6 we discuss the abundance
measurements, which in Section~7 are compared with the results of P05
and of other investigators.  Section~8 discusses the implications of the
photometric and the spectroscopic results on the SGB multiplicity. In
Section~9 we present relative age measurements for the spectroscopic
target stars.  A final section discusses the implications of the
observational facts presented in this paper for the stellar population
in $\omega$~Centauri and for the origin of this anomalous cluster.

\section{Photometry}

\subsection{HST data}

Our photometric \hst\ study is based on a mosaic of 3$\times$3 {\sl HST}
ACS/WFC fields taken in GO-9442 (PI Cool).  Each of the fields has
exposures of 3$\times$340 + 12 sec.\ in F435W, 3$\times$340 + 8
sec.\ in F625W, and 4$\times$440 sec.\ in F658N (H$_\alpha$).  The
images were reduced using {\sf img2xym\_WFC.09x10}, which is a publicly
available FORTRAN program described in Anderson \& King (2006).  The
program finds and measures each star in each exposure, by fitting a
spatially variable empirical PSF.  We collated the independent
measurements of the stars into a master star list that covers the entire
mosaic field.  For each star we constructed an average magnitude in each
band, and computed errors from the agreement among the independent
exposures.  The instrumental magnitudes were transformed into the ACS
Vega-mag flight system following the procedure given in Bedin et al.\
(2005), using the zero points of Sirianni et al.\ (2005).  To calculate
our aperture corrections we used single {\sf DRZ} images with the longer
exposure time.

\begin{figure*}[t!]
\centering
\includegraphics[width=14cm]{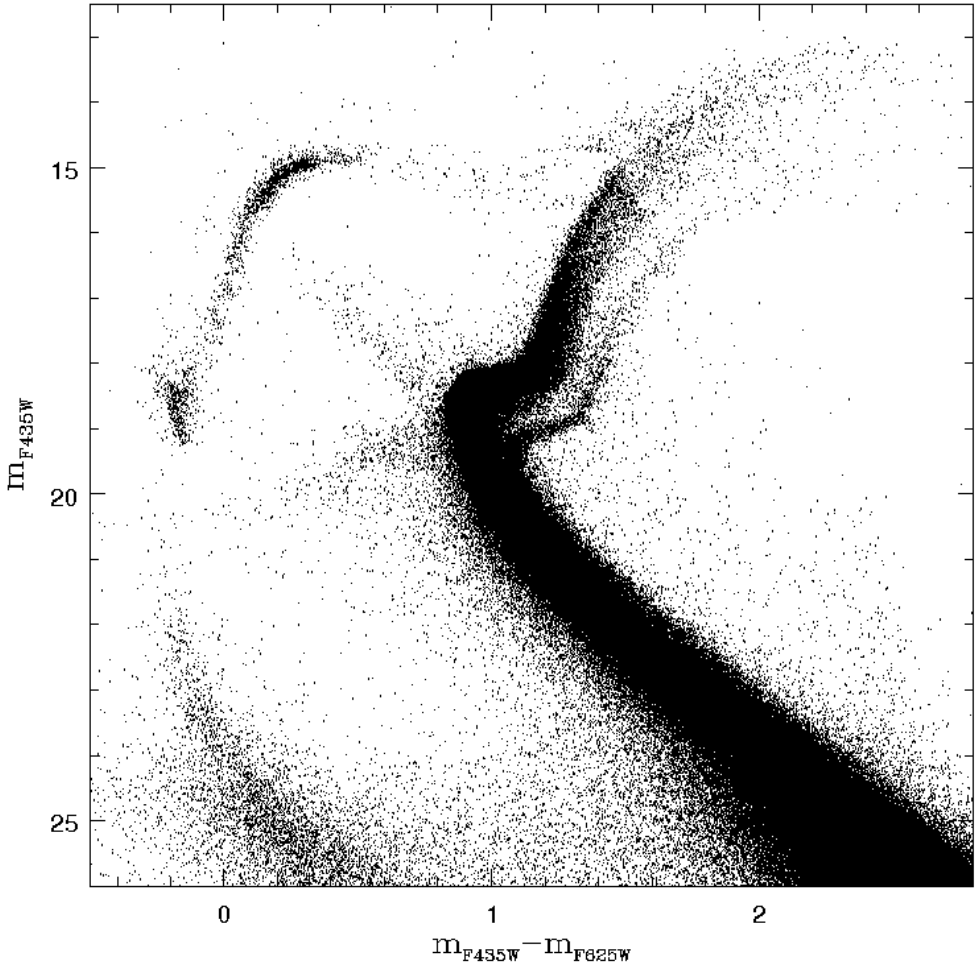}
\caption{Our CMD of $\omega$ Centauri, from
         the photometry of more than a million
         stars in the central $3\times3$ ACS fields.}
\label{f1}
\end{figure*}

Figure~\ref{f1} shows our color-magnitude diagram (CMD).  We believe
that this is the most accurate CMD ever published for this cluster.  In
the present paper we discuss the subgiant region and the upper part of
the main sequence; later papers will deal with other features.

\subsection{WFI data}

The ground-based component of our photometry is based on 187 images
taken from 1999 to 2003 with the WFI@ESO2.2m camera (hereafter WFI).
Exposure times cover the range 5--1800 seconds in Johnson $U$, $B$, and
$V$, and in Cousins $R_C$ and $I_C$, plus some observations
with the 665 nm and 658 nm narrow-band filters.  The images were reduced
using {\sf img2xym\_WFI}, software developed specifically for the WFI
camera and described in Anderson et al.\ (2006). The program works in a
way quite similar to {\sf img2xym\_WFC.09x10}, from which it was
derived.  We carried out photometry and astrometry of about 30,000
stars.
Taking advantage of the 4-year temporal baseline, we derived
proper motions, which, after application of corrections for differential
chromatic refraction, allowed a very good cluster/field star separation.

The whole WFI data set will be the subject of a separate paper.
For the present study we used the WFI data to choose the spectroscopic
targets and to derive their atmospheric parameters.

\section{Structure within regions of the CMD}

Our new CMDs provide us with three kinds of information.  First, they
allow us to delineate sequences better than we could before.  Second, in
several cases they show us connections between populations in different
regions of the CMD.  Third, when combined with spectroscopic abundances
for a modest number of stars, they will provide a crucial correspondence
between abundance-based populations and photometry-based ones.

\subsection{Main sequence}

\begin{figure*}[t!]
\centering
\includegraphics[width=14cm]{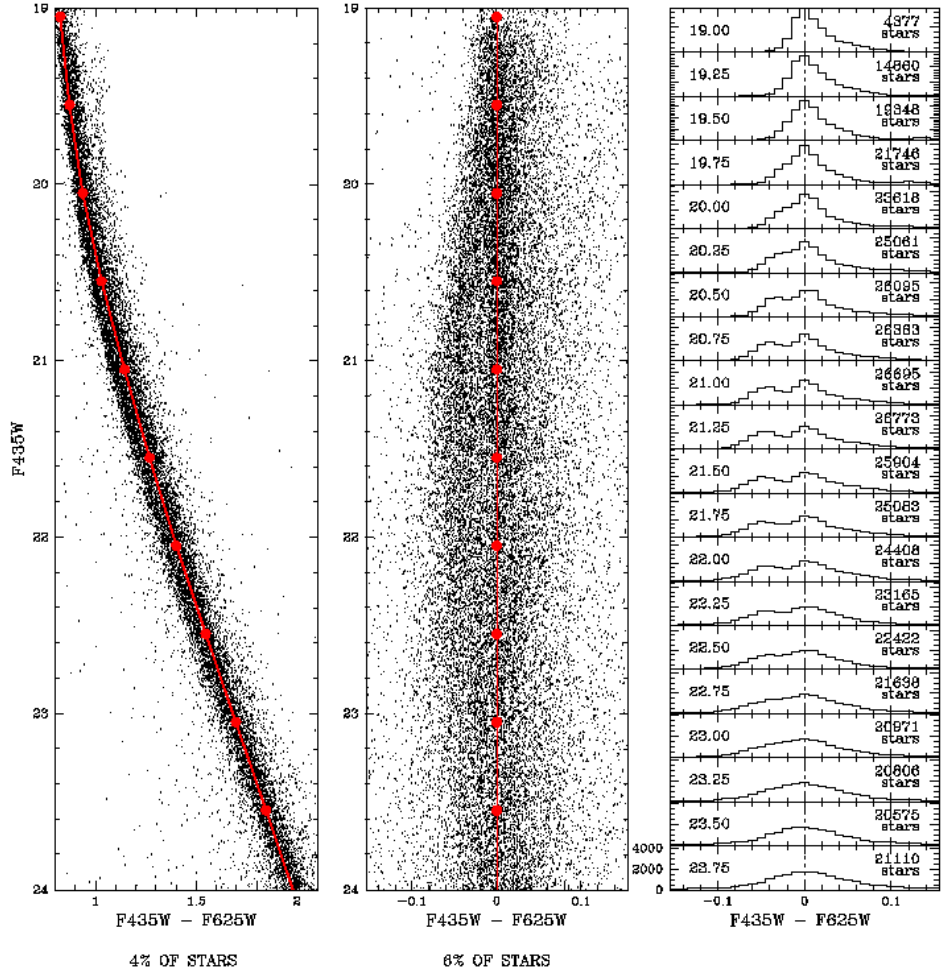}
\caption{Enlargement of the CMD of Fig.~\ref{f1}, showing the double main
   sequence of B04.  In the central panel we have subtracted from the
   color of the red MS the color of a fiducial line, drawn by hand.  The
   right-hand panel shows the color distribution of the points plotted
   in the central panel.}
\label{f2}
\end{figure*}

With our new more accurate photometry we can now study the double part
of the main sequence more carefully.  Since the split is not at all
clear in the MS area of Fig.~\ref{f1}, we plot in the left panel of
Figure~\ref{f2}
only a randomly chosen 4\% of the stars, while the middle panel shows a
randomly chosen 6\% of the stars on an expanded color scale, with a
hand-drawn fiducial sequence subtracted out.  These percentages were
chosen to reduce the number of stars to a level that would enable these
two plots to give a clear visual impression of the MS split.  Our
quantitative result is in the right-hand panel, which shows histograms
of the colors of all of the stars, in quarter-magnitude intervals.  The
distributions corresponding to the two sequences overlap, but they
clearly define two groups of stars, the first at $m_{\rm F435W}$ $-$
$m_{\rm F625W}$ $\sim$ 0 in the straightened figure and the second at
$m_{\rm F435W}$ $-$ $m_{\rm F625W}$ $\sim$ $-$0.05---the rMS and bMS of
B04, respectively---groups that are quite distinct, at least in the
magnitude interval 20.5 $<$ $m_{\rm F435W}$ $<$ 22.5.  The two MSs tend
to merge at brighter and fainter magnitudes.  No additional splits are
apparent within these two sequences
(See also the discussion in Sollima et al.\ 2006b.)

\begin{figure*}[t!]
\centering
\includegraphics[width=14cm]{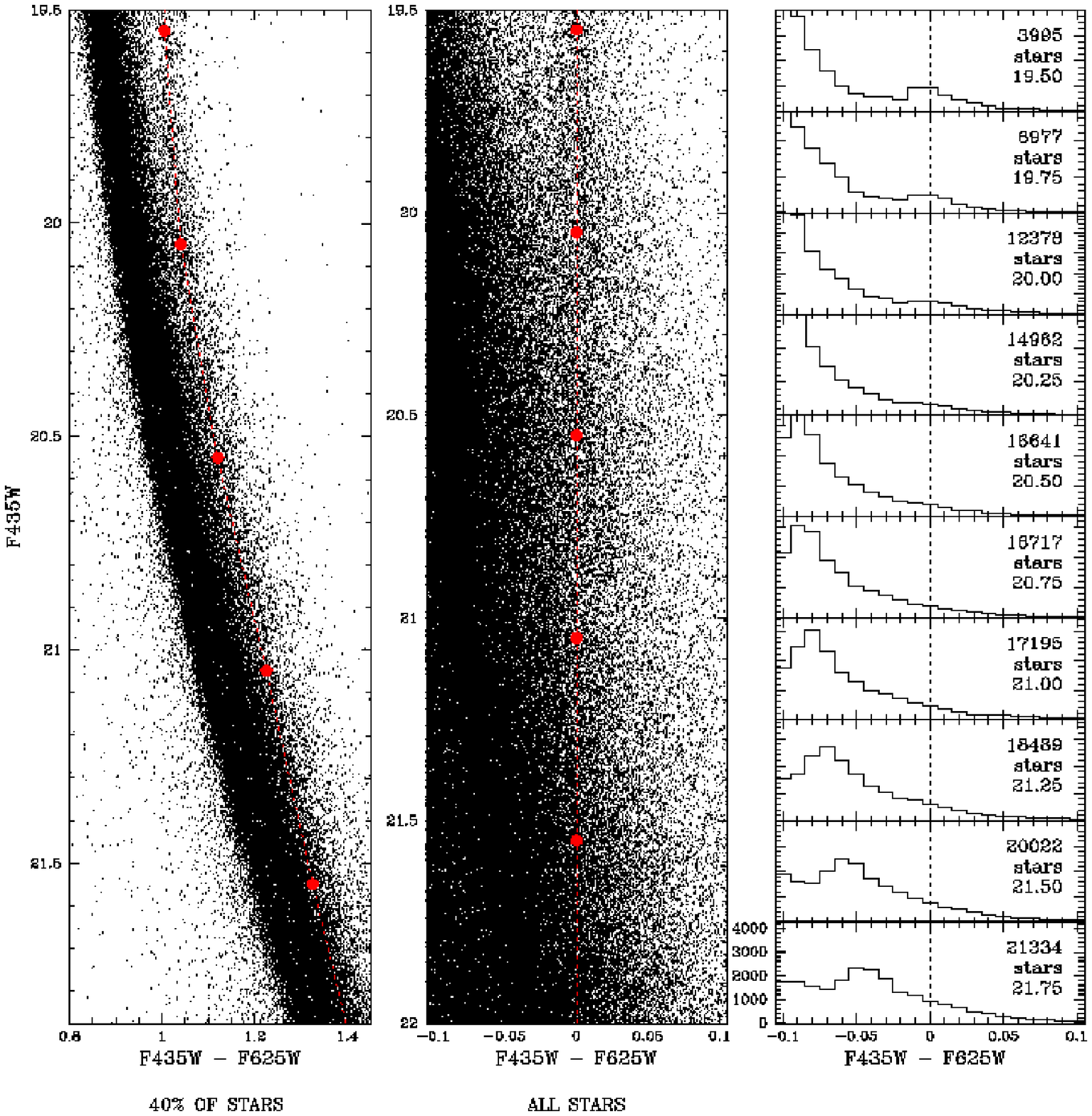}
\caption{The reddest branch of the main sequence, which connects with
   SGB-a.  In the left panel is an enlargement of part of the CMD, with
   a fiducial sequence drawn through the reddest branch.  In the middle
   panel the colors of the fiducial sequence are subtracted off, and at
   the right are histograms of 0.25-magnitude intervals.}
\label{f3}
\end{figure*}

We can also analyze another population aspect of the MS.  As was noted
by Sollima et al.\ (2005a, hereafter S05), in Fig.~\ref{f1} the detached
red-giant sequence RGB-a apparently continues through the subgiant
region (SGB-a) and then merges into the main sequence.  With our new
photometry we can follow it part way down the MS, as shown in
Figure~\ref{f3}.  In the histograms the sequence, whose main-sequence
turnoff in Fig.~\ref{f1} is at $B=19.4$, is visible as a secondary
maximum down to $B=20$, and appears to create an extension of the color
distribution down at least to $B=21$.  By analogy with RGB-a, we call
this MS-a.

In Fig.~\ref{f2} and Fig.~\ref{f3} we can see that where the bMS and
rMS are better separated, MS-a is blended with the rMS.  Because
of this blend, we can measure the fraction of stars belonging to the bMS
only with respect to the total MS population. We find that the
bMS contains $\sim$33\% of the main-sequence stars.  This value will
be useful in our later discussion.  Here we note that Sollima et
al.\ (2006b) have found a strong radial gradient for the ratio of
bMS/rMS stars, for distances $r>8$ arcmin from the cluster
center. Interestingly enough, while we confirm their external gradient
with independent {\sl HST} data, the radial distribution of the bMS/rMS
stars within our $3\times3$ ACS field is flat (Bellini et al., in
preparation). Therefore the fraction of bMS stars with respect to the
the other main sequence stars stays constant within
the {\sl HST} field analyzed in the present paper.

\subsection{Subgiant region}

We now turn to the subgiant region, which in the rest of this paper we
will use as a key to the multiple populations of $\omega$~Centauri.
Figure~\ref{f4} shows an enlargement of that part of the CMD.  This
region is even more complex than the main sequence.  In the lower part
of the figure is a histogram of the star numbers within the
parallelogram shown, summed parallel to the short edge and plotted
against a coordinate ($X$) that runs along the long edge, with zero
point at the heavy line in the upper panel.  Along with the four
distinct peaks labeled A, B, C, and D, there is a broad distribution of
stars in the interval $-0.8<X<-0.45$.  These populations are better
identified in the Hess diagram of the region that is plotted in
Fig.~\ref{f5}.  We will refer to them as SGB Groups A, B, C, and D.

Here we are facing important evidence that we will try to interpret in
later sections:\ the number distribution of the SGB stars across the CMD
is significantly different from what we would have expected from the
distribution of the stars across the branches of the main sequence.  The
rMS and bMS of B04 correspond to two distinct groups, separated in color
and with different metal and helium content (P05), and without any
apparent substructures (as shown in Fig.~\ref{f2}).  To these two
sequences we need to add MS-a, on the red side of the rMS, which we have
clearly identified in Fig.~\ref{f3}, and which apparently continues into
SGB Group D and then into RGB-a.  From the distribution of the stars on
the MS, we would have expected to see only three distinct SGBs, one
including all of the stars that have a metallicity similar to that of
the bMS, one with the stars that have the metallicity of the rMS, and a
third component coming from the reddest MS of Fig.~\ref{f3}, which is
identified with SGB Group D.  Contrary to these expectations, the SGB
region shown in Fig.~\ref{f5} is split into many different branches,
with stars distributed into a CMD region spanning 0.6--0.8 magnitude in
$m_{\rm F435W}$, if we exclude SGB Group D, or more than 1.2 magnitudes
if we include it.  It is important to note that this magnitude range is
more than twice as large as the 0.4--0.5 magnitude that would be
expected from the metallicity range covered by the $\omega$~Cen stars if
we assume that all of them have the same age, as can be seen from
isochrones published by Pietrinferni et al.\ (2004, 2006).

\subsection{Continuity between regions}

\begin{figure*}[ht!]
\epsscale{1.7}
\plotone{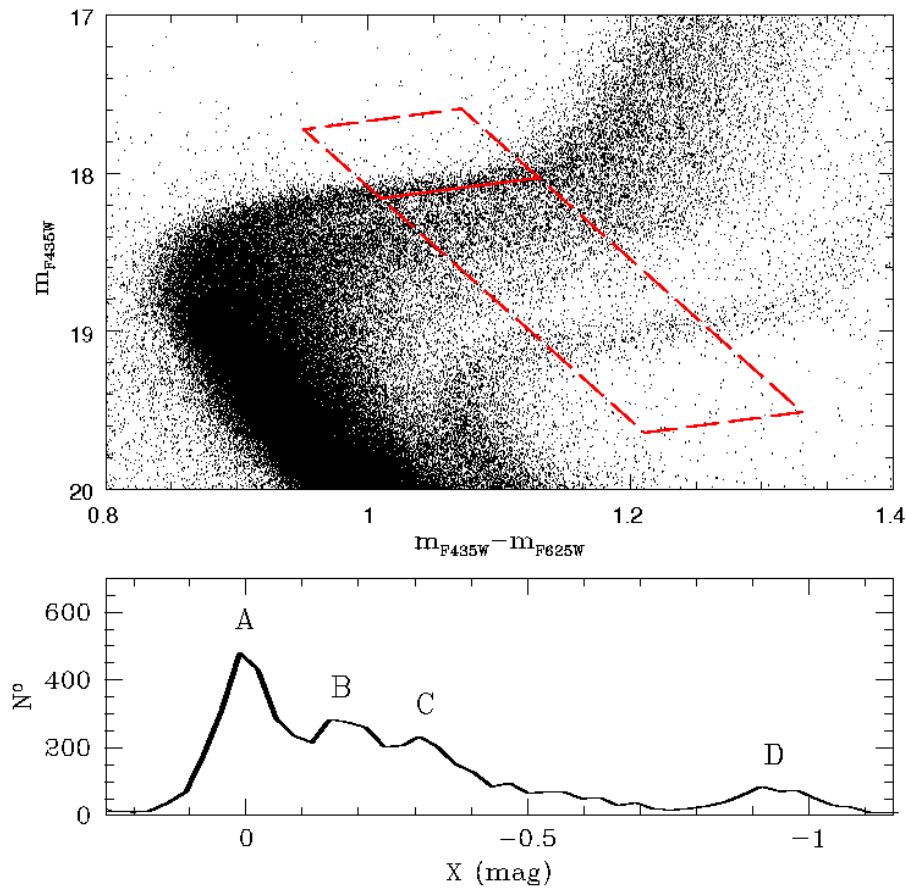}
\caption{Top:\ enlargement of Fig.~\ref{f1} to show the SGB region
         better.  Bottom:\ distribution of the stars in the
         parallelogram.  (See text for details.)}
\label{f4}
\end{figure*}

\begin{figure*}[ht!]
\epsscale{2.2}
\plotone{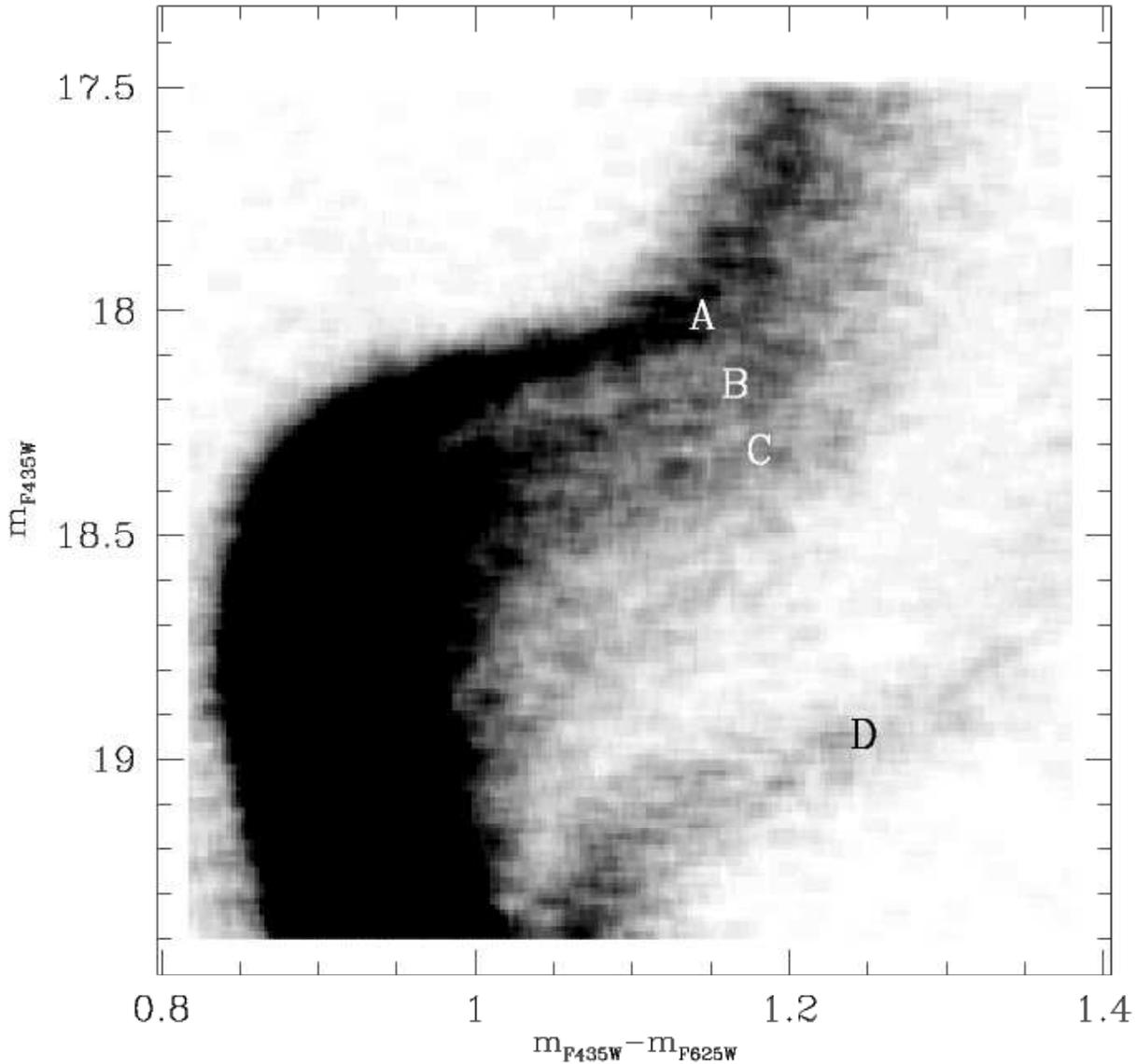}
\caption{Hess diagram of the SGB region,
  to show its complex structure. At least four
  agglomerates of stars can be distinguished (marked A, B, C, D), with
  additional stars distributed between the faintest group (SGB-a = D)
and the three brighter groups.}
\label{f5}
\end{figure*}

\begin{figure}[t!]
\epsscale{1.00}
\plotone{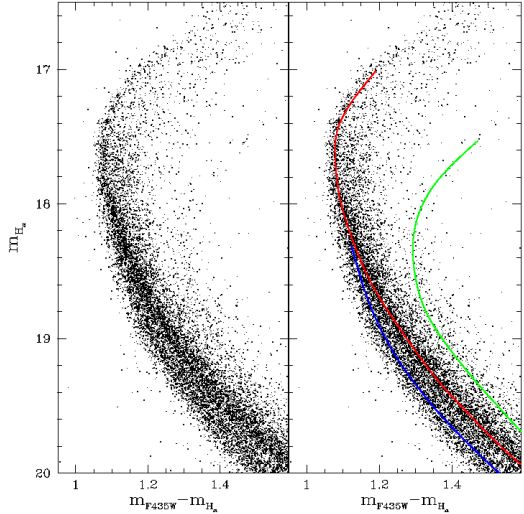}
\caption{(left) $m_{\rm H_{\alpha}}$ vs.\ $m_{\rm F435W} - m_{\rm
H_{\alpha}}$
  CMD for the ACS fields. (right) Fiducial lines drawn by
  hand help to show the connections between the MS and the SGB.
  It is clear that part of the rMS continues into SGB Group A
  (red line). Also MS-a is clearly connected with SGB
  Group D (green line).  By contrast, the bMS can be followed
  only up to the TO (blue line), then it spreads into the
  SGB region.}
\label{f6}
\end{figure}

The continuity of the sequences that correspond to the four SGB groups
is fairly clear within the SGB region, but it is not always obvious how
they connect with sequences in other parts of the CMD (both MS and RGB).
In particular, it is not at all clear how SGB Groups B and C connect
with the MS.
At this stage, however, we can make the following simple connections:

\begin{itemize}
\item
SGB Group D connects with the reddest branch of the MS, MS-a, as can
be seen from Figs.~\ref{f3}--\ref{f6}.
Figure~\ref{f6} is very
similar to panel {\it e} of Fig.~1 in B04, but to avoid confusion it
displays a randomly selected subsample of the stars in the ACS
fields. As in B04, here we use the $m_{\rm H_{\alpha}}$ vs.\ $m_{\rm
F435W}$-$m_{\rm H_{\alpha}}$ CMD in order better to show the
separation between the two branches in the upper part of the MS.

\item
It is clear that the rMS continues into the upper SGB
(which in the preceding subsection we named Group A),
as shown in Figure~\ref{f6}.

\item
The bMS can be followed up nearly to the turn-off, as can be seen in
Fig.~\ref{f6}.  It then continues into the SGB region, where it is no
longer possible to follow it clearly.

\end{itemize}

It is also clear that SGB Group A cannot account for all of the rMS
stars.  We estimated the rough number of stars in each group of
Figs.~\ref{f4} and~\ref{f5} by fitting four Gaussians centered on the
four peaks of the histogram of the lower panel of Fig.~\ref{f4} (see
Section 8).  SGB Group A contains about 33\% of the SGB stars, much less
than the fraction of MS stars that belong to the rMS (in our {\sl HST}
field, about two thirds).  We will resume this discussion at the end of
Section 8, where we will be able to combine this information with what
comes from the spectroscopic analysis.

%%%%%%%%%%%%%%%%%%%%%%%%%%%%%%
%
\section{Spectroscopic observations and data reduction}
%
%%%%%%%%%%%%%%%%%%%%%%%%%%%%%%

The spectroscopic data come from ESO DDT time [proposal 272.D-5065(A)],
and were collected in April--May 2004 with FLAMES@VLT\-+GIRAFFE. The sky
was clear, and the typical seeing was 0.8 arcsec.  We used the MEDUSA
mode, which obtains 132 spectra simultaneously.  To have enough
$S/N$, and in order to cover the wavelengths of interest, in the
spectrograph we used the LR2
set-up, which gives $R=6400$ in the 3960--4560 \AA \ range.
Thirty-four GIRAFFE fibers were placed on stars of the blue and red
branches of the MS; the results from those spectra have already been
presented in P05.  The remaining fibers were placed on SGB stars (88
fibers) and on the sky (10 fibers).
Twelve one-hour spectra were obtained for each target.

Twenty-two target stars were selected from the 3$\times$3 mosaic of {\sl
HST} fields presented in Section~2.1.  An additional 66 SGB stars were
selected from the ground-based $\sim 34' \times 33'$ ESO/WFI@2.2m field.
Its coverage includes the region of the {\sl HST} fields for which
photometric results were presented in Section~2.1.  Eight of the WFI
stars were eliminated from our list because their radial velocity was
not compatible with the cluster velocity and we therefore considered
them to be field stars.  The coordinates of our final targets are
reported in Col.~2 and 3 of Table~\ref{t1} (\hst\ sample), and
Table~\ref{t3} (WFI sample).  Note that the center of the cluster (which
can be used for radial distance determination of the the target stars)
is at $\alpha=201\fdg691208$, $\delta=-47\fdg476861$.

The data were reduced using GIRAFFE pipeline 1.12 (Blecha et al.\ 2000),
which corrects the spectra for bias and flat-field.  (See
http://girbldrs.sour\-ceforge.net/ for documentation on the GIRAFFE
pipeline and software.)  Then each spectrum was corrected for its fiber
transmission coefficient, which was found from five flat-field images,
by measuring for each fiber the average flux relative to a reference
fiber.  A sky correction was applied to each stellar spectrum by
subtracting the average of ten sky spectra that were observed
simultaneously with the stars (same FLAMES plate).  The wavelength
calibration uses both prior and simultaneous calibration-lamp spectra.
Finally, each spectrum was normalized to the continuum, i.e., divided by
a low-order polynomial that fits its continuum.  The resulting spectra
have a dispersion of 0.2 \AA/pixel and a typical $S/N\sim$ 100--150.

\begin{figure}[ht!]
\epsscale{1.0}
\plotone{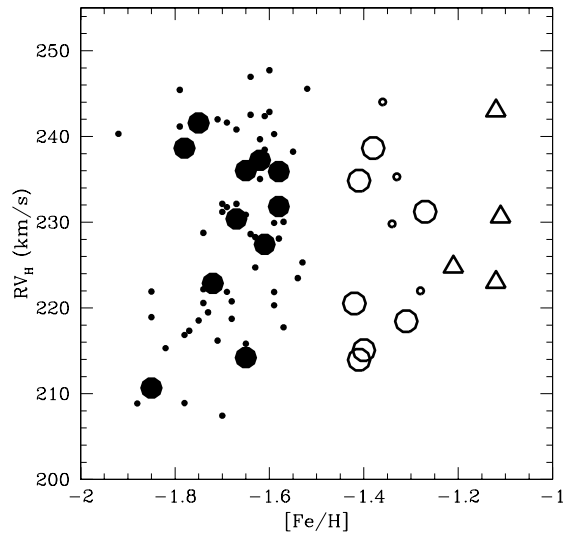}
\caption{Heliocentric radial velocities of the observed stars vs.\
          metallicity.  Large symbols are {\sl HST} stars, while small
          ones are WFI stars.  Open triangles are the stars with SGB-a
          metallicity.  Filled and open circles are SGB-MP and
          SGB-MInt2, respectively.  (See Sect.~5 for definitions of
          the metallicity groups.)}
\label{f7}
\end{figure}

We used the {\sf gyCrossC.py} utility of the GIRAFFE pipeline to
measure the radial velocity, which we then converted to heliocentric.
The resulting velocities for the member stars ($RV_{\rm{H}}$) are
given in Tables~\ref{t1} and~\ref{t3}, and shown in Fig.~\ref{f7},
where we plot radial velocity vs.\ metallicity.  (See Section~6 for
the metallicity determination.)  In this figure and in a number of
others, we distinguish between the \hst\ sample (large symbols) and
the WFI sample (small symbols), because they are at different
distances from the center of the cluster, and later studies might want
to make this distinction.
The error in radial velocity is typically about 2--3 km/s.  All the
stars have the same radial velocity within $\pm20$\ km/s.  Considering
the mean radial velocity of $\omega$~Cen ($\sim$232 km/s, Reijns et al.\
2006), the
velocity dispersion in the inner part of the cluster ($\sim$15 km/s,
Reijns et al.\ 2006) and the
observational errors, all of the stars in Fig.~7 appear to be cluster
members.

\begin{deluxetable}{rcccccccc}
\tabletypesize{\scriptsize}
\tablewidth{0pt}
\tablecaption{\scriptsize{\hst\  stars}}
\tablehead{
\colhead{ID} & \colhead{R.A.(J2000.0)} & \colhead{Decl.(J2000.0)} &
\colhead{$m_{\rm F435W}$} &  \colhead{$m_{\rm F435W}-m_{\rm F625W}$} &
\colhead{$V$} &
\colhead{$RV_{\rm{H}}$(km/s)} & \colhead{$T_{\rm{eff}}(K)$} &
\colhead{$\log g$}
}
\startdata
%\multicolumn{9}{c}{\small{SGB$-$MP $+$ SGB$-$MInt2 stars}}\\
3735  & 201.589019 &  $-$47.535167  & 18.08 & 1.07 & 17.38 & 214 & 5650 &
3.7\\
5533  & 201.584075 &  $-$47.488143  & 18.08 & 1.08 & 17.50 & 237 & 5670 &
3.8\\
7843  & 201.593368 &  $-$47.504161  & 18.06 & 1.09 & 17.39 & 236 & 5630 &
3.8\\
8756  & 201.607391 &  $-$47.560226  & 18.07 & 1.05 & 17.29 & 210 & 5680 &
3.8\\
13633 & 201.611114 &  $-$47.525005  & 18.42 & 1.12 & 17.63 & 238 & 5500 &
3.9\\
145   & 201.576721 &  $-$47.541572  & 18.04 & 1.08 & 17.26 & 222 & 5630 &
3.7\\
1472  & 201.565170 &  $-$47.455718  & 18.08 & 1.13 & 17.35 & 235 & 5540 &
3.7\\
2550  & 201.573867 &  $-$47.480480  & 18.32 & 1.15 & 17.64 & 238 & 5540 &
3.8\\
26656 & 201.641036 &  $-$47.551818  & 18.39 & 1.05 & 17.62 & 214 & 5760 &
4.0\\
299   & 201.557098 &  $-$47.442539  & 18.14 & 1.09 & 17.33 & 231 & 5680 &
3.7\\
3004  & 201.587936 &  $-$47.540699  & 18.36 & 1.04 & 17.66 & 227 & 5740 &
3.9\\
35208 & 201.653244 &  $-$47.554779  & 18.21 & 1.16 & 17.43 & 241 & 5430 &
3.8\\
3976  & 201.575729 &  $-$47.468921  & 18.31 & 1.10 & 17.68 & 218 & 5660 &
3.8\\
4079  & 201.595642 &  $-$47.561798  & 18.42 & 1.10 & 17.63 & 234 & 5660 &
4.0\\
4434  & 201.575195 &  $-$47.460380  & 18.26 & 1.04 & 17.65 & 215 & 5780 &
3.9\\
59481 & 201.680511 &  $-$47.552825  & 18.34 & 1.10 & 17.62 & 230 & 5600 &
3.9\\
63840 & 201.684097 &  $-$47.550704  & 18.29 & 1.09 & 17.60 & 231 & 5630 &
3.9\\
9462  & 201.608245 &  $-$47.556041  & 18.27 & 1.13 & 17.42 & 220 & 5580 &
3.8\\
%\multicolumn{9}{c}{\small{SGB$-$a stars}}\\
6808  & 201.593612 &  $-$47.517513  & 18.15 & 1.02 & 17.46 & 230 & 5840 &
3.9\\
28448 & 201.642211 &  $-$47.545017  & 19.00 & 1.24 & 18.22 & 243 & 5400 &
4.1\\
5654  & 201.596649 &  $-$47.545948  & 19.01 & 1.25 & 18.02 & 224 & 5350 &
4.0\\
6766  & 201.591690 &  $-$47.508903  & 18.97 & 1.16 & 18.24 & 223 & 5550 &
3.9\\
\enddata
\label{t1}
\end{deluxetable}

%
%%%%%%%%%%%%%%%%%%%%%%%%%%%%%%
%
\section{Terminology}
%
%%%%%%%%%%%%%%%%%%%%%%%%%%%%%%

At this point it is appropriate to consider what terminology we should
use to identify the different populations of $\omega$~Cen.

As the number of sequences recognized in the CMD of $\omega$~Cen has
increased, the number of designations for them has increased even
faster.
For the SGB, a useful terminology was introduced by Sollima et al.\
(2005a), who, by analogy with their
terminology for the RGB (Sollima et al.\ 2005b), referred to the most
metal-poor, most luminous of the subgiant branches as SGB-MP, the less
luminous, less metal-poor subgiant region, with two populations, as
SGB-Mint$n$ (where $n$ is 2 or 3), and the anomalous, apparently
detached, faintest subgiant branch as SGB-a.

It is tempting to continue to use the same suffixes to
identify the different SGBs, but the problem has become more
complicated, because a new distinction must now be made.  Whereas
previous authors have referred to populations either by their
metallicities or by locations in the color-magnitude diagram, in the
present paper we will be showing that these two types of population
criterion are not equivalent.  It is therefore important that our
terminology distinguish between them.  What we will do here is to
continue to refer to metallicity groups by the suffixes introduced by
Sollima et al., but to introduce also, where necessary, new terms to
distinguish populations according only to their locations in the CMD.

We will therefore use the suffixes -MP and -Mint2 only to designate
stars whose [Fe/H] values correspond approximately to those of the
corresponding
SGB metallicity groups of Sollima et al.\ (2005a).  (Our sample does not
happen to include any stars that belong to
the Sollima et al.\ SGB-Mint3 group.)
For sequences that have been identified photometrically, we use terms such
as ``SGB Group A''.  In the case of populations for which we use the
suffix -a, however, we do not make a distinction, because at present
neither their mean metallicity nor their metallicity dispersion is
clear.

When we use the labels A, B, C, and D for groups in the SGB region, we
should make it clear that these are arbitrary letters chosen for
convenience and are not meant to imply that these are the only branches
that will ever be distinguished.  We also wish to make it abundantly
clear that our designations within the CMD are meant for a particular
region, such as MS, SGB, or RGB, and that the connections between such
pieces are not necessarily known yet (see also Section~9 for a
discussion).

%%%%%%%%%%%%%%%%%%%%%%%%%%%%%%
%
\section{Abundance measurements}
%
%%%%%%%%%%%%%%%%%%%%%%%%%%%%%%

In this Section we derive the Fe, C, N, Ca, Ti, and Ba
abundances
for the 22 stars from the {\sl HST} fields, and the Fe
abundances for the 58 stars selected from the WFI field.  The
measurement of Fe for the WFI stars is important in order to have a
larger sample of stars for the estimates of relative ages that we
will make in Sect.\ 9.  The abundances of other elements for the WFI
stars, however, are of less importance
in the present paper, and will be presented in a future one.

For the stars in the {\sl HST} field (Table~\ref{t1}), we derived effective
temperatures ($T_{\rm{eff}}$) from the $m_{\rm F435W}$ $-$ $m_{\rm
F625W}$ color in the {\sl HST} CMD.  The relation between color and
$T_{\rm{eff}}$, as a function of [M/H] (by which we mean the global
metallicity, including alpha enhancement), was derived from isochrones
by Pietrinferni et al.\ (2004, 2006).  Colors were de-reddened using the
absorption coefficients listed in Table~3 of Bedin et al.\ (2005),
adopting $E(B-V)=0.115$.  As a first guess for the [M/H] to be used in
the color-[M/H]-temperature relation, we adopted [Fe/H] = $-$1.5, the
mean metallicity of $\omega$~Cen stars, along with an alpha enhancement
of 0.3 dex; this enhancement is confirmed {\it a posteriori} by our
abundance results.  The [M/H] was derived from the adopted [Fe/H] and
the alpha enhancement from the prescription by Salaris et al.\ (1993),
along with the corresponding $T_{\rm{eff}}$ from the
color-[M/H]-temperature relation.  Using this value for $T_{\rm{eff}}$,
we calculated $\log g$ and $v_{\rm{t}}$ and measured a new [Fe/H]
abundance as described below.  Then for each star the values of
$T_{\rm{eff}}$ and [Fe/H] were changed in an iterative process, till
convergence (when $\log g$ and $v_{\rm{t}}$ no longer change by a
significant amount).

We estimated the effect of variations in helium content and age on the
relation between color and temperature, as follows. Using the same set
of isochrones, we found that a variation $\Delta Y=\pm0.1$ in He
content, and a $\Delta$(age) = $\pm$2 Gyr each imply a variation of
$\sim$10 K in temperature for SGB stars, which translates into a change
of $\sim$0.01 dex in metallicity.  Such small changes can be neglected.

The gravity $\log g$ was calculated from the elementary formula
$$ \log\left(\frac{g}{g_{\odot}}\right) =
         \log\left(\frac{M}{M_{\odot}}\right)
         + 4 \log\left(\frac{T_{\rm{eff}}}{T_{\odot}}\right)
         - \log\left(\frac{L}{L_{\odot}}\right). $$
The mass $M/M_{\odot}$ was derived from the relations of Straizys \&
Kuriliene (1981) for the given $T_{\rm{eff}}$, adopting a luminosity
class IV for all the stars.  (Even though these relations were derived
for Population I stars, using them for Population II stars produces a
negligible error in gravity.)  The luminosity $L/L_{\odot}$ was derived
from the apparent magnitude $V$ measured in WFI images, assuming
the absolute distance modulus $(m-M)_0=13.75$ found by van de Ven et
al.\ (2006), and the reddening adopted above.  The bolometric
correction (BC) was derived from the BC-$T_{\rm{eff}}$ relation of
Alonso et al.\ (1999).  Finally, the microturbulent velocity came from
the relation (Houdashelt et al.\ 2000)
$$v_{\rm t} = 2.22 - 0.322 \log g .$$
For all the stars we found $v_{\rm t} \sim$ 1 km/s.

For the stars in the WFI field, effective temperatures were the mean
values of the temperatures derived from the color-[Fe/H]-temperature
relations of Alonso et al.\ (1996), Alonso et al.\ (1999), and Sekiguchi
\& Fukugita (2000), using the de-reddened $B-V$, $V-I_{\rm C}$, and
$V-R_{\rm C}$ colors.  The $\log g$ and $v_{\rm t}$ parameters were
obtained as for the {\sl HST} stars.  The adopted atmospheric
$T_{\rm{eff}}$ and $\log g$ are listed in Tables~\ref{t1} and~\ref{t3}
for the {\sl HST} and WFI stars, respectively.

Since our scales of effective temperature were derived in different ways
for the two different sets of targets (\hst\ and WFI), we verified that
they agree with each other.  To this purpose, we measured the
temperatures of the {\sl HST} stars following the same procedure that
was used for the WFI targets, i.e., using their WFI $B$, $V$, $R_{\rm
C}$, and $I_{\rm C}$ magnitudes, and the color-temperature relations of
Alonso et al.\ (1996), Alonso et al.\ (1999), and Sekiguchi \& Fukugita
(2000), in order to derive an independent value of $T_{\rm eff}$.  We
obtained temperatures which differ, on average, by less than 10 K from
the $T_{\rm eff}$ values that we had determined using the {\sl HST}
photometry and the Pietrinferni et al.\ isochrones.

The agreement between temperature scales implies that our adopted
color-temperature relations are consistent with each other.  We
therefore expect that systematic differences between our metal
abundances and those derived by others from these color-temperature
relations and Kurucz model atmospheres will be negligible.\\ On the
other hand, our absolute metal abundances could have systematic errors
of the order of 0.15--0.20 dex, because of systematic errors in
effective temperatures, and uncertainties in model atmospheres and in
reddening.

The metal content was obtained by comparison with synthetic spectra
calculated using SPECTRUM, the local-thermodynamical-equilibrium
spectral synthesis program freely distributed by Richard O.\ Gray.  (See
www.phys.app\-state.edu/spectrum/spectrum.html for more details.)  The
model atmospheres of Kurucz (1992), used throughout this paper, assume
$N_{\rm{He}}/N_{\rm{H}} = 0.1$, corresponding to $Y=0.28$ by mass.  The
bMS and, quite reasonably, the related SGB stars (i.e., stars with the
same metallicity as the bMS stars), were assumed to have a helium
content $Y\sim0.38$, in accordance with our assumption that the bMS
stars are helium-rich.  As discussed in P05, this increase in helium
introduces an error smaller than 0.03 dex in the metal-abundance
determinations, which is negligible.  We firstly measured a metallicity
index [A/H], as in P05. This index comes mainly from Fe lines, plus some
lines of $\alpha$-elements like Ca and Ti.  (See P05 for its exact
definition.)  We chose this index because it is the metallicity index
used by P05, and we wanted to compare the results of this paper with the
results of P05.  They had to resort to [A/H] (and not [Fe/H], as we will
do in the present paper) because of the lower $S/N$ of their spectra.
P05 called their metallicity index [M/H]; here we prefer to use a
different name in order to avoid confusion with the symbol [M/H] that is
commonly used for global metallicity, as we did above.

\begin{figure}[ht!]
\epsscale{1.0}
\plotone{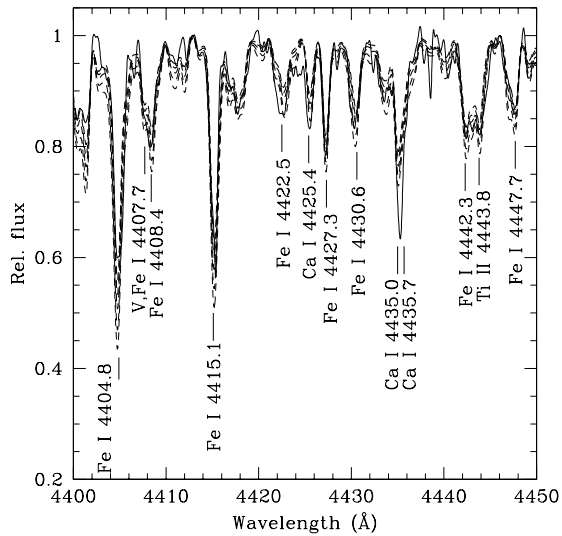}
\caption{The spectrum of star 28448 in the wavelength interval
         4400--4450 \AA\ (continuous line).  Superposed are the synthetic
         spectra for metallicities [Fe/H] $-1.4$, $-1.2$, $-1.0$,
         $-0.8$, $-0.6$ (dashed lines).  Many spectral lines are
         identified.}
\label{f8}
\end{figure}

Our [A/H] values were obtained from a comparison of each observed
spectrum with five synthetic ones (see Fig.~\ref{f8}), calculated with
different metal abundances but with other element ratios as in the Sun.
We normalized these spectra to the continuum, as we did for the observed
spectra.  The comparison was done in the 4400--4450 \AA\ interval,
because this region contains numerous metal lines (mainly due to Fe-peak
elements, with a few strong Ca and Ti lines) but has few lines due to
molecules (CH and CN), and no strong H lines.  The synthetic spectra
were smoothed to the resolution of the observed spectra.

\begin{deluxetable}{rccrccccc}
\tabletypesize{\scriptsize} \tablewidth{0pt}
\tablecaption{\scriptsize{Abundances and ages for the \hst\  stars.}}
\tablehead{ \colhead{ID} & \colhead{[A/H]} & \colhead{[Fe/H]} &
\colhead{[C/Fe]} & \colhead{[N/Fe]} & \colhead{[Ca/Fe]} &
\colhead{[Ti/Fe]} & \colhead{[Ba/Fe]} & \colhead{Age} }
\startdata
%\multicolumn{9}{c}{\small{SGB$-$MP $+$ SGB$-$MInt2 stars}}\\
3735  & $-$1.64 & $-$1.76 & 0.13 & 1.16 & 0.35 & 0.44 & 0.95 & 0.63 \\
5533  & $-$1.55 & $-$1.62 & 0.00 & 1.06 & 0.55 & 0.41 & 1.11 & 0.64 \\
7843  & $-$1.57 & $-$1.65 & 0.00 & 1.33 & 0.37 & 0.34 & 0.97 & 0.63 \\
8756  & $-$1.62 & $-$1.85 & 0.03 & 1.11 & 0.17 & 0.36 & 0.71 & 0.59 \\
13633 & $-$1.64 & $-$1.78 & 0.03 & 1.62 & 0.32 & 0.25 & 1.00 & 0.94 \\
145   & $-$1.61 & $-$1.72 & 0.06 & 1.49 & 0.43 & 0.37 & 0.62 & 0.61 \\
1472  & $-$1.54 & $-$1.58 & 0.35 & 1.07 & 0.33 & 0.24 & 0.73 & 0.69 \\
2550  & $-$1.27 & $-$1.38 & $-$0.05 & 1.48 & 0.42 & 0.43 & 0.88 & 0.79\\
26656 & $-$1.28 & $-$1.41 & $-$0.07 & 1.61 & 0.44 & 0.39 & 0.85 & 0.83\\
299   & $-$1.21 & $-$1.27 & 0.12 & 1.67 & 0.60 & 0.30 & 1.03 & 0.85 \\
3004  & $-$1.55 & $-$1.61 & 0.23 & 1.30 & 0.34 & 0.33 & 0.65 & 0.77 \\
35208 & $-$1.64 & $-$1.75 & 0.08 & 0.93 & 0.27 & 0.31 & 0.60 & 0.71 \\
3976  & $-$1.24 & $-$1.31 & $-$0.09 & 1.59 & 0.36 & 0.41 & 1.06 & 0.59\\
4079  & $-$1.29 & $-$1.41 & $-$0.13 & 1.67 & 0.26 & 0.36 & 1.00 & 0.72\\
4434  & $-$1.27 & $-$1.40 & 0.00 & 1.73 & 0.32 & 0.28 & 1.01 & 0.78 \\
59481 & $-$1.57 & $-$1.67 & $-$0.01 & 1.58 & 0.42 & 0.37 & 0.82 & 0.62\\
63840 & $-$1.31 & $-$1.48 & $-$0.20 & 1.55 & 0.33 & 0.26 & 0.99 & 0.68\\
9462  & $-$1.32 & $-$1.42 & $-$0.22 & 1.44 & 0.43 & 0.24 & 0.85 & 0.72\\
%\multicolumn{9}{c}{\small{SGB$-$a stars}}\\
6808  & $-$1.03 & $-$1.11 & $-$0.18 & 1.72 & 0.41 & 0.49 & 0.94 & 0.64\\
28448 & $-$1.02 & $-$1.12 & $-$0.18 & 1.73 & 0.55 & 0.30 & 0.99 & 1.01\\
5654  & $-$1.05 & $-$1.21 & 0.00 & 1.70 & 0.46 & 0.56 & 1.03 & 1.01\\
6766  & $-$1.01 & $-$1.12 & 0.03 & 1.67 & 0.52 & 0.41 & 1.06 & 0.98 \\
\enddata
\label{t2}
\end{deluxetable}

\begin{figure}[ht!]
\epsscale{1.0}
\plotone{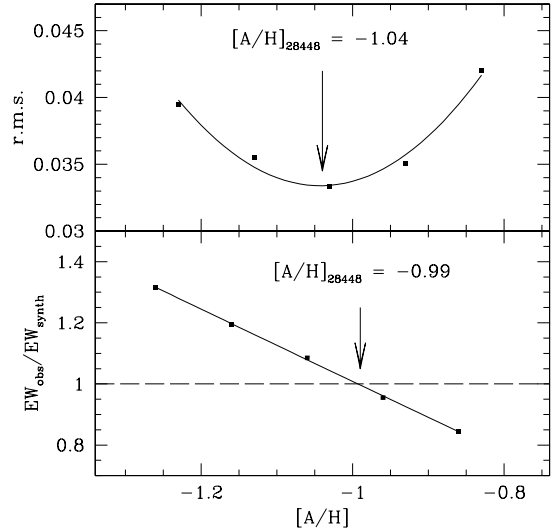}
\caption{The r.m.s.\ scatter (upper panel), and the empirical relation
         between [A/H] and EQW$_{\rm{obs}}$/EQW$_{\rm{synth}}$ (lower
         panel), for star 24448.  The metal abundance of the star is
         given by the minimum value of the r.m.s.\ curve (first method)
         and by the [Fe/H] value corresponding to
         EQW$_{\rm{obs}}$/EQW$_{\rm{synth}}$ = 1 (second method).}
\label{f9}
\end{figure}

The metallicity was obtained by two different methods.  The first one is
the method used by P05:\ their [A/H] is the value that minimizes the
r.m.s.\ scatter of the differences between the observed and synthetic
spectra (as illustrated in the upper panel of Fig.~\ref{f9}).  In the
second method, we measured an equivalent width (EW) for the whole
4400--4450 \AA\ region.  The EW of the observed spectrum
(EW$_{\rm{obs}}$) was then divided by the EWs of the theoretical ones
(EW$_{\rm{synth}}$), and an empirical relation was derived between
EW$_{\rm{obs}}$/EW$_{\rm{synth}}$ and [A/H] (as shown in the lower panel
of Fig.~\ref{f9}).  The resulting [A/H] is the value that corresponds to
EW$_{\rm{obs}}$/EW$_{\rm{synth}}$ = 1, as obtained from interpolation in
the empirical relation.  We have verified that the metallicities
obtained in these two ways agree within 0.05 dex,
allowing us to estimate that the error due to the methods is
less than 0.03 dex.  The final adopted metallicity
is the average of the values from the two methods.  As a test of our
methodology, we used a solar spectrum from the ESO archive
(http://archive.eso.org/), obtained with the same instrument and the
same configuration (FLAMES@VLT+GIRAFFE, LR2 mode), and we calculated a
synthetic spectrum for the Sun using the canonical solar atmospheric
parameters ($T_{\rm{eff}}=5777\ K$, $\log g=4.44$, $v_{\rm{t}}=0.8$).
We verified that our procedure (i.e., the program, the model atmosphere,
and the line list used) reproduces the strengths of the solar spectral
features.
For this purpose we measured the abundances of the Sun for the elements
considered in this paper by the same method used for the target
stars. We obtained:\ [Fe/H] = $-0.02$, [C/Fe] = $-0.10$, [N/Fe] = $+0.08$,
[Ca/Fe] = $+0.05$, [Ti/Fe] = $-0.07$, [Ba/Fe] = $+0.10$. We conclude that
our procedure reproduces the solar values well, within 0.10 dex.

For the measurement of the [Fe/H] value we applied the same methods, but
restricted the comparison of the observed and theoretical spectra to the
4400--4425 \AA\ region, which contains only iron lines. The smaller
spectral interval, with fewer spectral lines, implies a larger error in
the final metallicity.  The mean difference between [A/H] and [Fe/H] is
0.11 dex (useful for comparing the results of this paper with P05), with
[A/H] higher than [Fe/H], as expected because of the
$\alpha$-enhancement.  The [A/H] and [Fe/H] values derived for the 22
{\sl HST} stars are listed in Table~\ref{t2}, while Table~\ref{t3} gives
the [Fe/H]
abundances obtained for the 58 WFI stars.  As explained above, no
measurement of [A/H] was made for the WFI stars.

We found the accidental error in [A/H] from the distribution of
differences between the values derived from equal halves of our
wavelength range; typical errors for the whole range are 0.04--0.05 dex.
Typical errors for [Fe/H] should be 0.06--0.07, since [Fe/H] is obtained
from a spectral region only half as long.  To these errors we should add
(in quadrature) the error due to photometric uncertainty in the colors;
the error in color is typically of the order of 0.01 magnitude, which
translates into a 0.02 dex error in abundance.  After allowing also for
the error that comes from the uncertainty in the He content (less than
0.03 dex, as we have shown in P05), we adopt an overall
uncertainty of 0.06 dex for [A/H] and 0.08 dex for [Fe/H].  This is the
internal error in our metallicity measurement.  In addition there can be
a systematic error of the order of 0.15--0.20 dex, because of systematic
uncertainties in effective temperatures, and uncertainties in model
atmospheres and in reddening. The systematic errors do not affect the
relative metallicities of the different stellar populations of
$\omega$~Cen that we will discuss in later sections.

As a final test, we considered the possible contamination of the spectra
by close neighboring stars.  This problem might affect the metallicity
measurement of the \hst\ targets, which are all near the rather crowded
center; all WFI targets are located in much less crowded outer regions.
In the {\sl HST} images, we measured instrumental magnitudes of the target
stars and the neighbors, and their separations.
In order to minimize contamination, the choice of stars for targeting
had not allowed any star with a neighbor that is closer than 2.4 arcsec
(twice the diameter of a fiber) and is fainter by less than 2 magnitudes.
There is, in fact, one neighbor 0.6 arcsec from a target star, but it is
$\sim$2.5 magnitudes fainter.  For this worst case we calculated synthetic
spectra for the neighbor and for the target star, assuming as worst-case
metallicities [Fe/H] = $-1.7$ for the target star and [Fe/H] = $-1.1$
for the neighbor.
We summed the two spectra, weighting for the magnitude difference and
for the flux captured by the fiber, and derived a metallicity [A/H]; the
result differed by less than 0.03 dex from the metallicity of the target
star.

\begin{figure}[ht!]
\epsscale{1.0}
\plotone{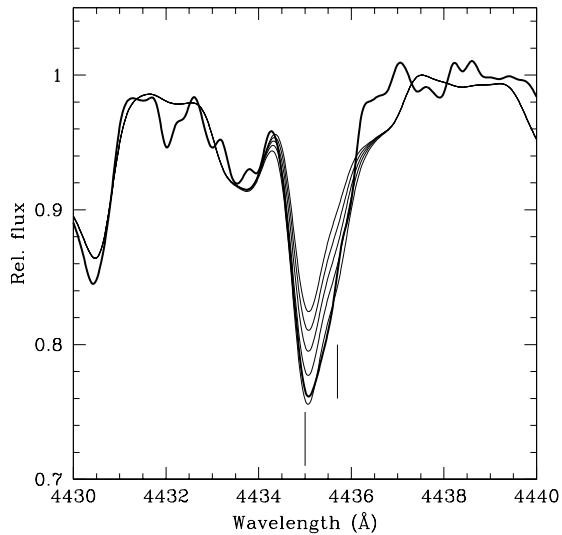}
\caption{The spectrum of the star 6766, compared with synthetic spectra
         in the region 4430--4440 \AA, which includes the Ca I lines at
         4435.0 and 4435.7 \AA.  (Their locations are marked.)  Synthetic
         spectra were computed for Ca abundances [Ca/Fe] = $-$0.2,
         0.0, +0.2, +0.4, +0.6 dex.  Thick line is the observed
         spectrum; thin lines are the synthetic spectra.  }
\label{f10}
\end{figure}

\begin{figure}[hb!]
\epsscale{1.0}
\plotone{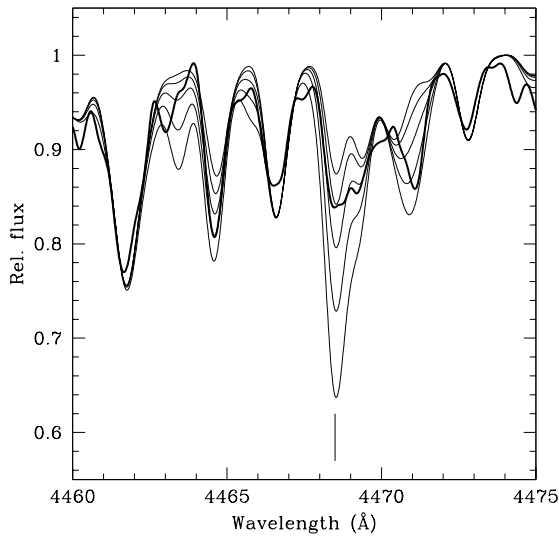}
\caption{The spectrum of the star 28448, compared with synthetic spectra
         in the region 4460--4475 \AA, which includes the Ti II line at
         4468.5 \AA.  (The location of the line is marked.)  Synthetic
         spectra were computed for Ti abundances [Ti/Fe] = $-$0.6,
         $-$0.2, +0.2, +0.6, +1.0 dex.  Thick line is the observed
         spectrum; thin lines are the synthetic spectra.  }
\label{f11}
\end{figure}

\begin{figure}[ht!]
\epsscale{1.0}
\plotone{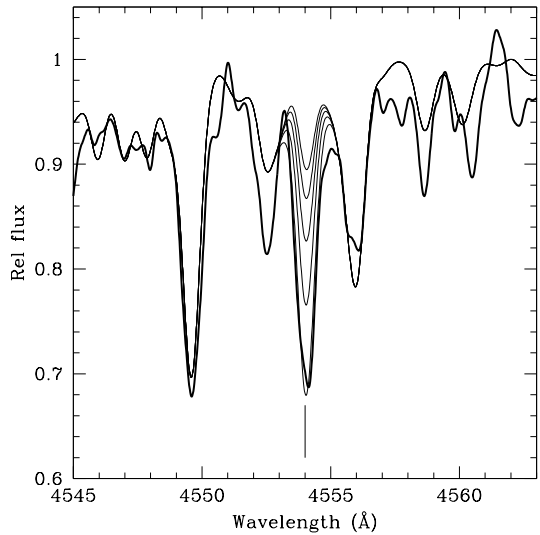}
\caption{The spectrum of the star 5654, compared with synthetic spectra
         in the region 4545--4563 \AA, which includes the Ba II resonance
         line at 4554 \AA.  (The location of the line is marked.)
         Synthetic spectra were computed for Ba abundances [Ba/Fe] =
         $-$0.6, $-$0.2, +0.2, +0.6, +1.0 dex.  Thick line is the
         observed spectrum; thin lines are the synthetic spectra.  }
\label{f12}
\end{figure}

\begin{figure}[hb!]
\epsscale{1.0}
\plotone{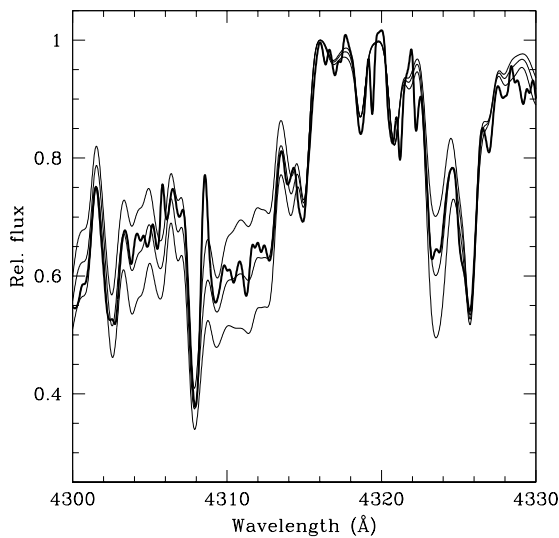}
\caption{The spectrum of the star 5654, compared with synthetic spectra
         in the region 4300--4330 \AA, which includes the CH band head.
         Synthetic spectra were computed for C abundances [C/Fe] =
         $-$0.2, 0.0, +0.2 dex.  Thick line is the observed spectrum;
         thin lines are the synthetic spectra.  }
\label{f13}
\end{figure}

\begin{figure}[ht!]
\epsscale{1.0}
\plotone{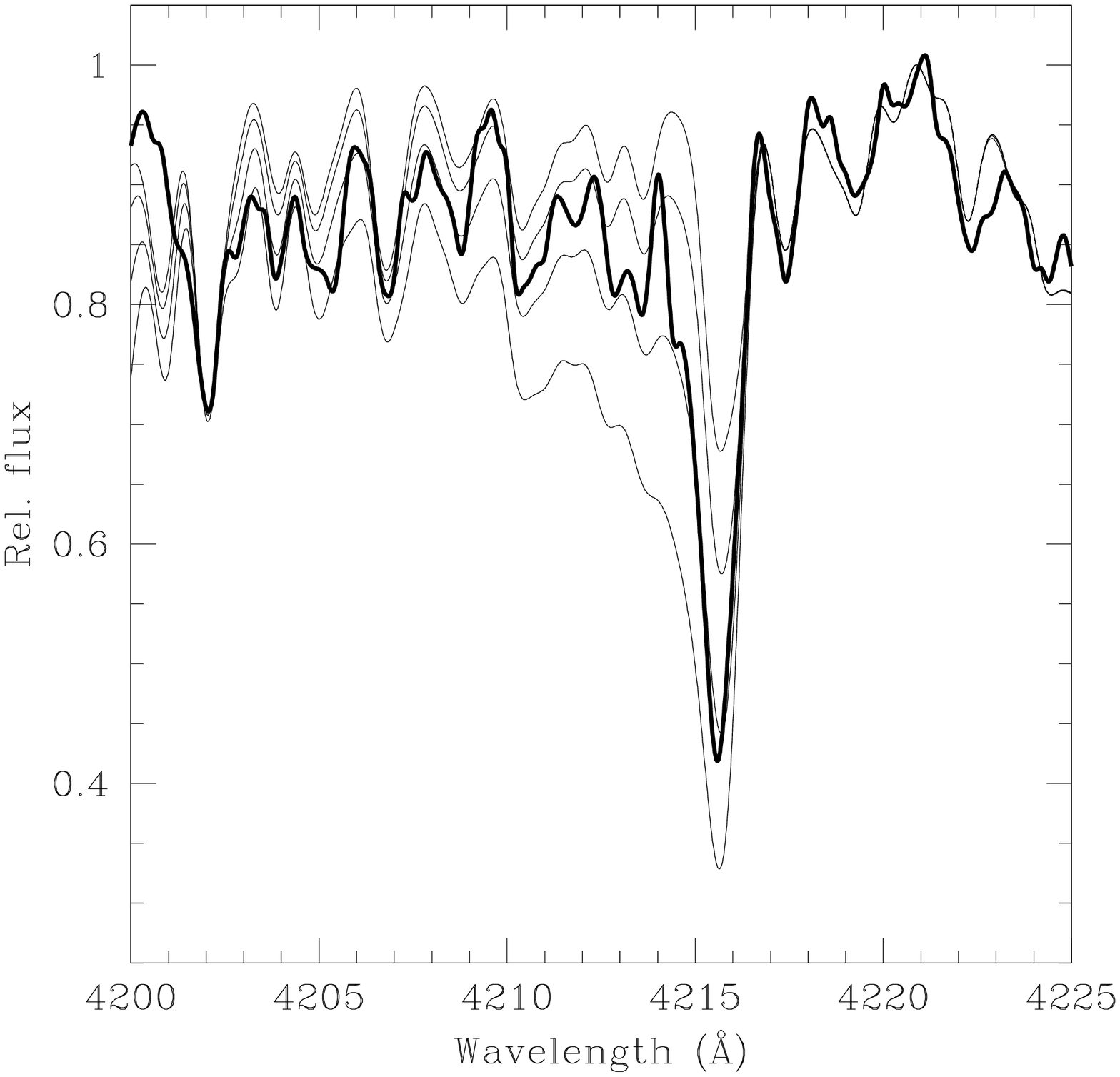}
\caption{The spectrum of the star 6766, compared with synthetic
         spectra in the region 4200--4225 \AA, which includes the band
         heads of the $\Delta v = 2$--0 violet CN band.  The synthetic
         spectra were computed for appropriate C content and for N
         abundances [N/Fe] = +1.0, +1.4, +1.8, +2.0 dex.  Thick line is
         observed spectrum; thin lines are synthetic spectra.}
\label{f14}
\end{figure}

In addition to [Fe/H], we were also interested in the abundances of
other elements.  We have been able to measure the abundances of C, N,
Ca, Ti, and Ba for the 22 {\sl HST} stars.  Calcium, titanium, and
barium abundances were obtained from the spectral lines of Ca I at 4435
\AA\ (Fig.~\ref{f10}), Ti II at 4468 \AA\ (Fig.~\ref{f11}), and Ba II at
4454 \AA\ (Fig.~\ref{f12}).  Carbon abundances were obtained by
comparing the observed spectra with synthetic ones in the spectral
region 4300--4330 \AA\ (see Fig.~\ref{f13}), which includes the $\Delta v =
0$ strong band heads of the A $^2\Delta$ -- X $^2\Pi$ transition of CH,
computed with appropriate model-atmosphere parameters, and different
values of the C abundances.  Nitrogen abundances were found by a similar
comparison for the region 4200--4225 \AA\ (see Fig.~\ref{f14}),
which includes the $\Delta v = -1$ band heads of the X $^2\Sigma$ -- B
$^2\Sigma$ CN transition.  All the abundances are listed in Table 2.

An upper limit for the errors in C, N, Ca, Ti, and Ba content was
calculated by assuming that the abundances of these elements are the
same within the intermediate-metallicity population (SGB-MInt2) and
within the metal-rich population (SGB-a), and calculating the average
r.m.s.\ spread of the measured abundance distributions of these five
elements.  (See Sect.~5 for definitions of these populations.)  The
final error is $\sim$0.1 dex.  This error should be considered an
overestimate of the uncertainties in the C, N, Ca, Ti, and Ba
abundances, because of the possibility of an intrinsic dispersion for
these elements among the measured stars.  In any case, we cannot do
better, because the abundances were obtained from a spectral interval
that is too small (too few lines) for us to apply the compare-two-halves
method that was used to estimate the uncertainty in [A/H].

\begin{figure}[hb!]
\epsscale{1.0}
\plotone{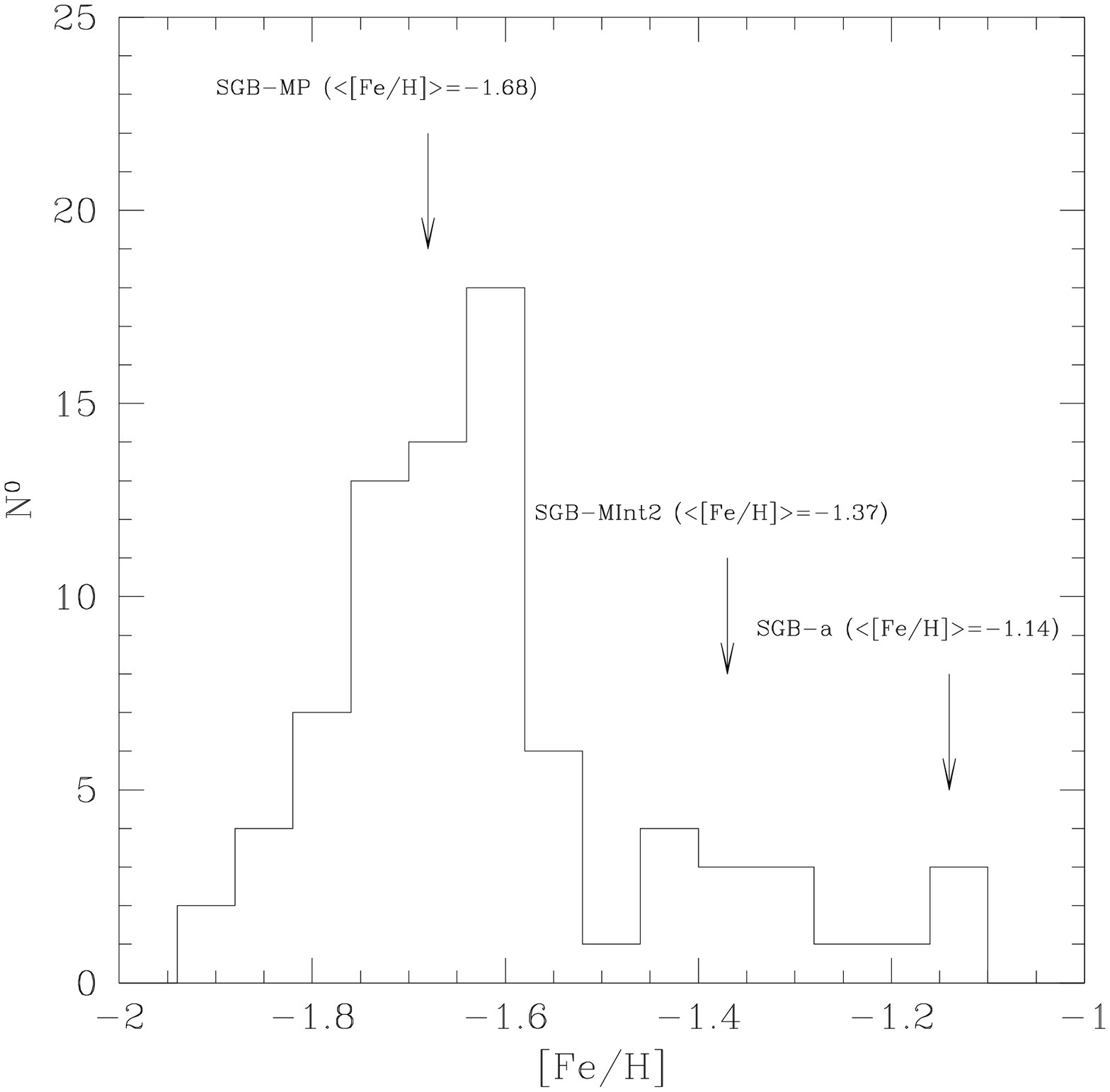}
\caption {Histogram of the metallicity distribution of the target stars.
          Three peaks are present. The first (SGB-MP), at
          $\rm{[Fe/H]}\sim-1.7$, has the same average metallicity as the
          rMS of B04, and corresponds to the bulk of the RGB
          population. The second one (SGB-MInt2), at
          $\rm{[Fe/H]}\sim-1.4$, has the same average metallicity as the
          bMS. The third one, at $\rm{[Fe/H]}\sim-1.1$, corresponds to
          the SGB-a population.}
\label{f15}
\end{figure}

%
%###################
%
\section{Discussion of the spectroscopic results}
%
%###################

Figure~\ref{f15} shows the histogram of the [Fe/H] distribution of all our
target stars.  (Because of the small number of stars observed, and the
selection biases, the figure cannot be construed as showing the
quantitative
metallicity distribution of $\omega$~Cen stars, but it does give the
shape of the distribution).  In the figure we can distinguish three
groups of stars, whose average metallicities are in agreement with three
of the four metallicity groups in the recent spectroscopic survey of SGB
stars in $\omega$~Cen by S05, who found four populations running through
the SGB region.  We can identify our peak at $\rm{[Fe/H]}\sim-1.7$ with
their SGB-MP and our peak at $\rm{[Fe/H]}\sim-1.4$ with their SGB-MInt2.
[We omit the Mint1 of the Sollima et al.\ (2005b) photometric paper
that deals only with the RGB,
as it does not appear in their spectroscopic paper (2005a = S05)
on the SGB.
Also, our sample does not include any stars in the part of the CMD where
they identify a fourth component, which they call SGB-MInt3.]  Our peak
at $\rm{[Fe/H]}\sim-1.1$ can be identified with their SGB-a.

The SGB-a group needs some further comments.  There is no doubt that as
defined by S05 (see their Fig.~1), this is the faintest SGB, corresponding
to our SGB Group D.  Therefore SGB-a is well identified, at least from
the photometric point of view.  However, it is not at all clear why S05
assign a metallicity [Fe/H] = $-0.6$ to this group of stars. Indeed,
panel $d$ of their Fig.~4 shows that the SGB-a stars in their sample
have a double-peaked metallicity distribution, with one peak at [Fe/H]
$\sim-1.0$ (as confirmed by our observations), and a second one at
[Fe/H] $\sim-0.6$.  Unlike S05, in our survey we have not identified any
stars with [Fe/H] $\sim-0.6$, but their absence may be due to the small
number
of SGB-a stars that we observed.  On the other hand, Pancino et al.\
(2002) found $\langle[\rm{Fe/H}]\rangle\sim-0.6 \pm 0.15$ from
high-resolution spectroscopy of three stars on RGB-a (which appears to
be the continuation of SGB-a on the RGB), confirming the
presence in this sequence of a population with a metallicity as high as
[Fe/H] = $-0.6$.

The dispersion of the stars of SGB-a in the CMD (as shown in
Figs.~\ref{f4} and~\ref{f5}) is clearly not consistent with as broad a
dispersion in the metal content of the SGB-a stars as was found by S05.
To show this we compared two isochrones from Pietrinferni et al.\ (2004,
2006), having the same age but with [Fe/H] = $-1.1$ and [Fe/H] = $-0.6$,
respectively.  Their separation in magnitude is more than twice as large
as the magnitude dispersion actually observed in SGB-a.  One possible
explanation for the large metallicity spread in the SGB-a sample of S05
is that their [Fe/H] estimates are based on the Ca II triplet.  Such
measurements are subject to large errors, because the relation between
the equivalent width of the Ca II triplet and [Fe/H] depends strongly on
the gravity (which corresponds to luminosity).  S05 noted that they
needed to calibrate this relation, but the catalog of reference
spectra that they used includes very few stars in the appropriate
ranges of temperature and gravity.  Their SGB-a sample spans more than a
magnitude in $R$; it is not out of the question that such a large
magnitude spread could be at least partially responsible for the large
metallicity dispersion that they find for the SGB-a stars.\\
To complicate the scenario further, the referee has noted that Norris \&
Da Costa (1995) found a group of RGB stars with [Fe/H]$\sim-1.1$, which
are apparently not present in RGB-a.  This means
that in the RGB region there are two groups  of stars, one with
[Fe/H]$\sim-1.1$, and a second with
[Fe/H]$\sim-0.6$. The former could be the progeny of our SGB-a
stars.\\
In short, we believe that the problem of the metal content of the
SGB-a stars and of a possible dispersion in it, and the problem of the
connection of the SGB-a sequence with the RGB-a of Pancino et al.\
(2000),
and with the RGB in general, are both still open, and require
further investigation.

\subsection{Comparison with P05}

In order better to understand the connection between the multiple SGBs
in $\omega$~Cen and the multiple MSs, it is useful to compare the
results presented in this paper with those of P05.  There it was found
that [A/H]
$=-$1.57 for the rMS and [A/H] $=-$1.26 for the bMS; when we apply the
0.11 dex correction needed to transform [A/H] into [Fe/H], as described
in Sect.~6, these values correspond respectively to [Fe/H] $=-$1.68 and
[Fe/H] $=-$1.37.  In the present paper we found for the SGB-MP stars a
mean [Fe/H] of $-$1.68 dex, and for the SGB-MInt2 stars a mean [Fe/H] of
$-$1.37 dex.  At this point, on the basis of the metallicity
measurements available, it is very tempting to identify the SGB-MP stars
as the progeny of the rMS stars, and the SGB-MInt2 stars as the progeny
of the bMS stars.  We will return to this question in later sections.

\begin{figure*}[t!]
\centering
\includegraphics[width=15cm]{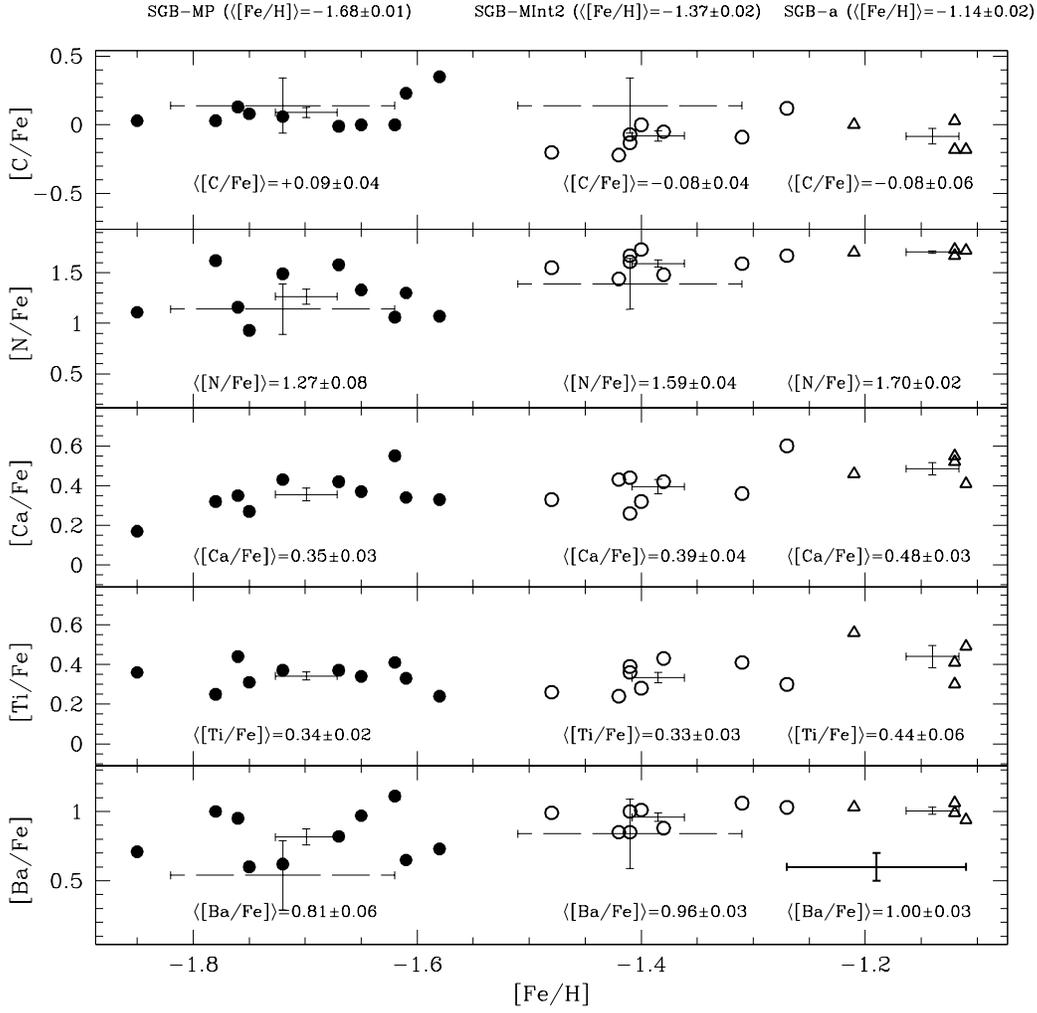}
\caption{The C, N, Ca, Ti, and Ba abundances for the 22 {\sl HST} stars.
         For each population the mean metallicity ([Fe/H]) is labeled at
         the top of the figure.  For each element and each population we
         labeled and plotted the mean value of the abundances
         (continuous lines), with the error of the mean, and the results
         of P05, when available (dashed lines). The continuous thick
         line at the lower right shows the error for a single measure.
         }
\label{f16}
\end{figure*}

In P05 we determined the abundances of C, N, and Ba.  Here, in addition
to those we have extended our analysis to include the $\alpha$-elements
Ca and Ti.  Figure~\ref{f16} shows the trend of the abundance ratios as
a function of [Fe/H].  Our measured abundances of C, N, and Ba for SGB
stars are in good agreement with the results of P05 for the MS.  In P05,
for both branches of the MS we found [C/A] $\sim0$, corresponding to
[C/Fe] $\sim0.1$ after application of the 0.11 dex correction to
transform the [A/H] into [Fe/H], as discussed in Section~6; here, for
metal-poor stars (the filled circles in Fig.~\ref{f16}) [C/Fe] spreads
from $-$0.1 to +0.4 dex, while for intermediate-metallicity stars (the
open circles in Fig.~\ref{f16}) [C/Fe] runs from $-$0.3 to 0.1 dex.  As
for the nitrogen content, in P05 we found for the rMS
[N/A] $\leq$ 1.0 ([N/Fe]
$\leq $1.1
after applying the 0.11 dex correction), and [N/A] $\sim$
1.0--1.5 ([N/Fe] $\sim$ 1.1--1.6) for the bMS; here, for metal-poor stars
(corresponding to the rMS) [N/Fe] spreads over a range from +0.9 to +1.6
dex, while for intermediate-metallicity stars (corresponding to the bMS)
N spreads over a range from +1.4 to +1.7 dex.
In summary, for low metallicities ([Fe/H] $\sim - 1.7$) the mean
abundances of C and N are [C/Fe] $\sim$ 0.1 and [N/Fe] $\sim$ 1.3,
respectively, while when the metallicity increases [C/Fe] decreases to
$\sim$ $-$0.1 and [N/Fe] increases to $\sim$1.6--1.7.  Compared with the
MS stars, SGB stars have slightly lower C abundance, and higher N
abundance.

Concerning Ba, in P05 we found [Ba/A] $\sim$ 0.4 ([Ba/Fe] $\sim$ 0.5)
for the rMS and [Ba/A] $\sim$ 0.7 ([Ba/Fe]$\sim$ 0.8) for the bMS; here,
for metal-poor stars [Ba/Fe] spreads over a range from +0.5 to +1.1 dex,
while for intermediate-metallicity stars [Ba/Fe] spreads over a range
from +0.8 to +1.1
dex.  We note that metal-poor stars have a Ba content lower than
intermediate-metallicity ones by about 0.2 dex.

Finally, we analyzed the alpha elements.  Both Ca and Ti show, in both
metal-poor and interme\-diate-metallicity stars, the enhancement of about
0.3--0.4 dex that is typical of intermediate-metallici\-ty globular
clusters.  SGB-a (the open triangles in Fig.~\ref{f16}) shows a chemical
composition similar to the SGB-MInt2 populations,
with a marginally greater (0.1 dex) excess of alpha elements:\
[C/Fe] $\sim -0.1$ , [N/Fe] $\sim
+1.7$, [Ca/Fe] $\sim$ +0.5, [Ti/Fe] $\sim$ +0.4, and [Ba/Fe] $\sim$
+1.0.

As a final point, it is worth noting that the C and N content of the
SGB-MInt2 population (as well as of the metal-rich one) is consistent
with what is predicted by Maeder \& Meynet (2006) for star-forming gas
contaminated by the ejecta of fast-rotating metal-poor massive stars;
most importantly, this could explain the very high He yield implied by
the results of P05 (see Maeder \& Meynet 2006 for further details).

\subsection{Comparison with other investigations}

The metallicity groups identified in Fig.~\ref{f15} correspond to similar
groups already identified on the RGB of $\omega$~Cen, and the average
metallicity of all our SGB stars is in general agreement with the
results of Suntzeff \& Kraft (1996), Norris et al.\ (1996), Pancino et
al.\ (2000), and Rey et al.\ (2004).  In fact, these previous studies
show the presence of at least three peaks in the metallicity
distribution of the RGB stars of $\omega$~Cen, the first at
$\rm{[Fe/H]}\sim-1.7$ (corresponding our SGB-MP group), the second at
$\rm{[Fe/H]}\sim-1.4$ (corresponding to our SGB-MInt2), and the third at
$\rm{[Fe/H]}\sim-1.0$ (corresponding to our SGB-a).

It is useful to compare our results for the abundances of C, N, Ca, Ti,
and Ba in SGB stars with those of Norris \& Da Costa (1995, hereafter
ND95) for the RGB stars, as ND95 present the most complete and extended
results on the chemical composition of $\omega$~Cen.  ND95 found a large
spread in [C/Fe], which spans the interval from $-$0.8 to +0.2 dex; the
spread is clearly present in the entire metallicity range ($-1.8<$
[Fe/H] $<-0.8$) covered by the stars of their sample.  We found $-0.2<$
[C/Fe] $<0.4$, with a tendency for the most metal-poor stars to have a C
content higher by $\sim$0.2 dex.  The offset between the C abundances
obtained in this paper and those of ND95 can be explained by the
different evolutionary phase, since it is well known that stars brighter
than the RGB bump (as are those observed by ND95) are depleted in C with
respect to subgiants (see Gratton et al.\ 2000).  As for the N content,
the agreement is less satisfactory, though we must take note of the
large errors (0.5 dex) of the ND95 measurements of N abundances.  ND95
found a large spread in [N/Fe], spanning the interval from 0.0 to 1.0
dex at every metallicity.  On average, our [N/Fe] values are larger by
$\sim$0.6 dex.
(Also, ND95 notice that their mean [N/Fe] value is 0.4--0.5
lower than in other work, e.g., Brown \& Wallerstein 1993.)\\
We also find a large dispersion in [N/Fe] for the SGB-MP,
while at higher metallicity we have [N/Fe] $\sim 1.6$ with a small
dispersion.  There is good agreement between our study and ND95 when we
consider Ca and Ti (both of them alpha elements).  Both the present
paper and ND95 find for the stars of $\omega$~Cen the alpha-element
overabundance of about 0.3--0.4 dex that is typical of intermediate and
metal-poor globular clusters.  There is agreement between the Ba content
measured in the present work and that of ND95, for the SGB-MP stars
(with a small overabundance, by 0.1--0.2 dex, in our results), but at
higher metallicity we have an overabundance of $\sim$0.3--0.4 dex with
respect to ND95.

Another important contribution to knowledge of the chemical composition
of $\omega$~Cen was made by Smith et al.\ (2000). In this case, our
results agree well:\ Smith et al.\ found a Ca and Ti content between 0.2
and 0.4 dex, as we do; but, more interesting, the Ba abundance also
agrees, with the bulk of stars in their paper having [Ba/Fe] between 0.5
and 1.0 dex.

%##########################
%
\section{The structure of the subgiant branch}
%
%##########################

At this point it is very instructive to put together the results from
the photometric and the spectroscopic investigations, in order better
to characterize the structure of the SGB of $\omega$~Cen, the
distribution of the stars along it, the properties of the stars in the
different SGB groups, and the connection of the SGB sequences with those
in the MS and RGB regions.

On the basis of the results of P05 on the metal content of the two MSs
in $\omega$~Cen, and of the metallicity measurements presented in the
present paper, it seems appropriate to identify the SGB stars in the
group that has [Fe/H] $\sim$ $-$1.7 (SGB-MP) as the progeny of the rMS
stars, and the SGB population that has [Fe/H] $\sim$ $-$1.4 (SGB-MInt2)
as the progeny of the bMS stars, as discussed in Section~7.1.

In order to complete the observational scenario, Figs.~\ref{f17}
and~\ref{f18} show the location in the CMD of the SGB stars whose metal
content we have measured.  Figure~\ref{f17} is limited to the \hst\
spectroscopic sample, and shows the CMD from the \hst\ data.
Figure~\ref{f18} shows all of the target stars, in a CMD that combines
\hst\ and WFI data.  Note that our \hst\ and WFI photometries have
nearly the same accuracy; although in uncrowded regions our \hst\
photometry tends to be more accurate than ground-based photometry, the
\hst\ fields used here are in the crowded central part of the cluster.
Even so, the much larger \hst\ sample gives better separation of the
different SGB groups.  From Figs.~\ref{f17} and~\ref{f18} we can
conclude that:

\begin{itemize}

\item All of the stars in our sample that come from SGB Group A (which
  peaks at $X=0$ in Fig.~\ref{f4}) are metal poor and fall within the
  SGB-MP metallicity group.  As discussed in Section~7.1, the SGB-MP
  stars have the same metal content as the rMS stars. As suggested by
  the CMD of Fig.~\ref{f6}, the spectroscopic results confirm that the
  SGB stars of group A are probably the progeny of the rMS.

\item The SGB-MP stars ($-1.90\le$ [Fe/H] $\le-1.50$) show a large
  magnitude dispersion, $\sim0.5$ mag in $m_{\rm F435W}$ or $B$.

\item None of our intermediate-metallicity SGB-MInt2 stars ($-1.50\leq$
  [Fe/H] $\le-1.25$) is located on SGB Group A; apparently, SGB-MInt2
  stars are fainter ($X<-0.1$) than SGB Group A.

\item The SGB-MInt2 stars also show a large dispersion in magnitude,
  though a smaller one than the SGB-MP stars:\ $\sim$0.3 magnitude in
  $m_{\rm F435W}$ or $B$.  As discussed in Section~7.1, the SGB-MInt2
  stars have the same metal content as the bMS stars, and therefore
  at least part of them are likely to be the progeny of that sequence.

\item The stars in the faintest SGB group, D (the peak at $X=-0.95$ in
  Fig.~\ref{f4}), all have [Fe/H] $\sim-1.1$.

\item None of the stars in our sample, not even the three stars in
  SGB group D, have [Fe/H] $\sim-0.6$.

\item There is one star (No.\ 6808) that has [Fe/H] = $-1.1$ (and
  therefore belongs to the SGB-a metallicity group) but is located in
  the uppermost SGB group, A, in the CMD.  It is marked in
  Fig.~\ref{f17}.  It is well separated from any neighbors in the ACS
  field, and its photometric measurements pass all of the selection
  criteria that identify the stars with the best photometry in the ACS
  catalog.  We consider its location in the CMD to be well established.

\item Unfortunately, our sample contains no stars in the range
  $-0.80<X<-0.45$ in Fig.~\ref{f4} (which should correspond to the
  SGB-MInt3 group of S05).

\end{itemize}

With the help of the histogram in the lower panel of Fig.~\ref{f4}, and
taking into account the conclusions just stated, we can make a rough
estimate of the fraction of stars in each of the SGB populations.  We
began by fitting four Gaussians to the distribution of SGB stars shown
in the lower panel of Fig.~\ref{f4}, centered at the four peaks marked
in the histogram.  We then integrated the areas under these Gaussians,
and found that SGB Group A includes $\sim$33\% of the total stars, SGB
Group B $\sim$29\%, SGB Group C $\sim$20\%, and SGB Group D (which
corresponds to SGB-a) $\sim$10\% of the total SGB population.  The
remaining $\sim$8\% is distributed in the region corresponding to
$-0.80<X<-0.45$.

\section{The Subpopulations of $\omega$ Cen}
%
%%%%%%%%%%%%%%%%%%%%%%%%%%%%%%

Before trying to understand the formation history of this intriguing
cluster, we need to identify and characterize each of its discrete
subpopulations. Individual subpopulations can be identified either
photometrically, from position in the CMD, or spectroscopically, by
metallicity.  To complicate the issue, while such identifications can be
made separately on the main branches of the CMD (MS, SGB, RGB, HB),
for most cases in $\omega$ Centauri it is not obvious from
the CMD alone how a population identified on a specific branch connects
tonthose on other branches, and, in particular, which are the MS and RGB
partners of a given SGB population.

On the MS we have three well-identified populations (see
Figs.~\ref{f2} and \ref{f3}), which in Section 3.1 we called rMS, bMS, and
MS-a.  On the SGB there are four different sequences (see
Figs.~\ref{f4} and \ref{f5}), which in Section 3.2 we called SGB Groups A,
B, C, and D, and there is a hint of a fifth sequence at a position
intermediate between Groups C and D.  The RGB has a very broad
component, and a much redder component that is clearly detached.
According to Sollima et al.\ (2005b) the RGB splits into five
subcomponents, which they call RGB-MP, RGB-Mint1, RGB-Mint2,
RGB-Mint3, and RGB-a, in order of increasing metallicity.  While
RGB-MP and RGB-a are well-established components, in the CMD of
Fig.~\ref{f1} we cannot clearly distinguish the three RGB-MInt
populations,
partly because stars in the upper part of the RGB are saturated in our
images, and this degrades the quality of our photometry.
Therefore we cannot directly confirm the significance of all of the
components proposed by Sollima et al.\ (2005b), despite the fact that
our CMD includes a larger sample of stars
because of our somewhat larger area.

The SGB offers one unique opportunity for identifying subpopulations.
On the main sequence, stars that have the same composition occupy the
same position regardless of age, so that different ages cannot be
distinguished, while on the RGB the substructure is controlled mainly by
the metallicity distribution.  Thus only the SGB can be used to estimate
relative ages of the various subpopulations.

Several questions naturally arise from this empirical taxonomy of the
various CMD branches: 1) How does each of the 5 SGB components map into
the MS and RGB components?  2) What are the age, metallicity, and helium
content of each subpopulation?
3) What fraction of the total number of stars belongs to each
subpopulation?

\subsection{Identifying evolutionary connections between MS, SGB, and
  RGB}

The first question can be addressed using three complementary kinds of
evidence:\ a) the morphological continuity from one branch in the CMD
to another, b) the relative numbers of stars on the various branches,
and c) the spectroscopic metallicities of individual stars on the
various branches.  The information that is available is summarized in
Table~\ref{t4}.  Each column refers to a different section of the CMD
and lists the groups that we recognize there.  For each group we give
the name,
the percentage of the stars that belong to that group (with each column
adding up to 100\%), and the metallicity [Fe/H].
In Table~4, we indicate more than one [Fe/H] value if it appears that
one group identified on the CMD may include stars belonging to
different metallicity groups.
MS and SGB percentages
and metallicities are from the present study and P05.  For the RGB we
have adopted the nomenclature, percentages, and metallicities from
Sollima et al.\ (2005b), assuming [Fe/H] = [M/H] $-$ 0.3.

An inspection of the various CMDs (e.g., as in Fig.~6) allows us to
identify only two quite obvious connections, namely rMS $\rightarrow$ SGB
Group A $\rightarrow$ RGB-MP, and MS-a $\rightarrow$ SGB Group
D $\rightarrow$ RGB-a. Further insight can come from
considering the relative percentages of the groups in the different
branches.  Note, however, that one does not expect the relative
percentages of a given population to be exactly the same in all regions
of the CMD.  In fact, on the MS a given magnitude interval corresponds
to different stellar mass intervals for different compositions, and the
evolutionary rates on the SGB and RGB are also a function of composition.
However, these effects are typically at the level of 10--20\% (cf.\
Fig. 13 in Zoccali et al.\ 2003) and will be ignored here.

In spite of the clear morphological connection rMS $\rightarrow\ $SGB
Group A, this cannot be the whole story, because the percentages do not
match, as is shown clearly in the first of the five sections of Table 4.
The rMS makes up a much larger fraction of the MS than Group A does of
the SGB; many of its stars must therefore evolve into other parts of the
SGB.  Since our spectroscopic investigation found that SGB Groups B and
C also include stars with [Fe/H] $\sim$ $-$1.7, we suggest that the
other rMS stars evolve into these two groups, although there is no
indication of the proportions.  But in any case, all of the metal-poor
stars with [Fe/H] $\sim -1.7$ must eventually connect to the the RGB-MP
population, whose percentage is not totally inconsistent with that of
the rMS group.

We now come to the helium-rich bMS population.
Combining the photometric and spectroscopic information of
Fig.~\ref{f6} and of Figs.~\ref{f17} and \ref{f18}, we conclude that
the bMS must continue through SGB Groups B and C, though it is not
clear in what proportion.

The subsample of SGB Group B and Group C stars with [Fe/H] $\sim -1.4$
(which, given their metallicity, do not evolve into RGB-MP)
must continue through RGB-MInt, as, presumably, must the stars between
Groups C and D too.
But it is not clear in what proportion they occupy
the different RGB-MInt branches proposed by Sollima et al.\ (2005b).

Finally, we note an apparent inconsistency between the percentages of
SGB Group D and RGB-a; the former includes $\sim$10\% of the SGB stars,
whereas the latter has only $\sim$5\% of the RGB stars, although a tight
connection between the two is apparent from the morphology of the CMD.
However, we also find some inconsistency between our metallicity for
SGB Group D, [Fe/H] = $-$1.1, and the metallicity found by
Pancino et al.\ (2002), [Fe/H] = $-$0.6
(an ambiguity that we also include in Table~4).

\subsection{The relative ages of the subpopulations}

The empirical facts that emerge from the present photometric and
spectroscopic investigation can be summarized as follows:

\begin{itemize}

\item The stars that populate the SGB of $\omega$~Cen cover a large
    magnitude interval, up to $\sim1.2$ in $m_{\rm F435W}$ or $B$. Four
    distinct SGB groups can be clearly identified, plus a possible
    fifth
    group that contributes a smaller percentage.  This contrasts with
    what one would have expected from the MS alone, where only three
    distinct stellar groups can be identified.

\item SGB stars in the most metal-poor group ([Fe/H] $\sim -1.7$)
have a 0.3--0.4-magnitude spread in luminosity
at a given color.  The intermediate-metallicity group ([Fe/H] $\sim
-1.4$) also shows some evidence of a similar range in SGB luminosity at
a given color.

\item The most metal-rich stars ([Fe/H] $\sim-1.1$) populate
   the faintest SGB, with a small dispersion in magnitude; however, one of
   them is 0.85 magnitude brighter in $m_{\rm F435W}$ than the others.

\end{itemize}

The most straightforward explanation for the large magnitude
dispersion of the metal-poor SGB stars on the flat part of the SGB is
in terms of an age difference.  An age difference is also indicated by
the luminosity difference among intermediate-metallicity stars that
populate SGB Groups B and C, though we know that the issue is
complicated by the fact that more than half of the stars in these
branches must be the progeny of the helium-rich bMS population.  So
for these branches we cannot firmly distinguish age effects from
helium effects.  It is important to note here that a spread in age has
a negligible influence on the MS position, and its influence on the
location of the RGB is also small.  On the other hand, it has a great
influence on the slope and location of the SGB in the CMD; however,
the SGB location is also affected by the helium abundance.

Taking advantage of our accurate photometry and of the metallicity
measurements from our spectra, we can now try to estimate the relative
ages of the individual SGB stars.  We do not attempt absolute age
determinations here, however, because they would be sensitive to the
assumed distance and reddening of the cluster, and to the photometric
zero points that we have adopted.  All that we assume is that the
distance and the reddening are the same for all of the stars in the
cluster.

\begin{figure}[h!]
\epsscale{1.0}
\plotone{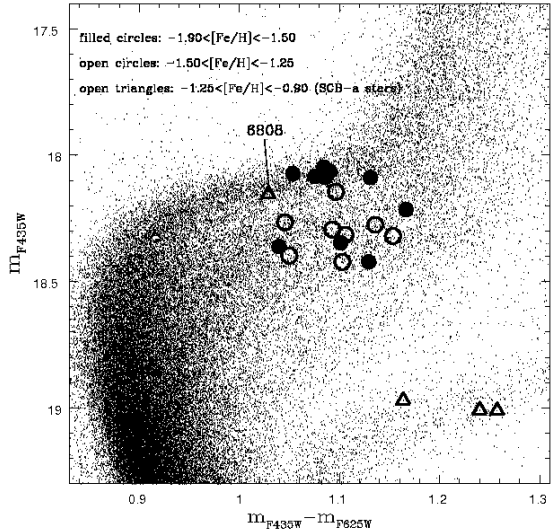}
\caption{Locations of the 22 \hst\ stars in the ACS CMD.  Different
          symbols indicate different metallicity groups, as indicated in
          the internal label.  The anomalous star 6808 is indicated.
          Note, here and in Fig.~\ref{f18}, the poor correlation between
          metallicity and location in the CMD.}
\label{f17}
\end{figure}

\begin{figure}[h!]
\epsscale{1.0}
\plotone{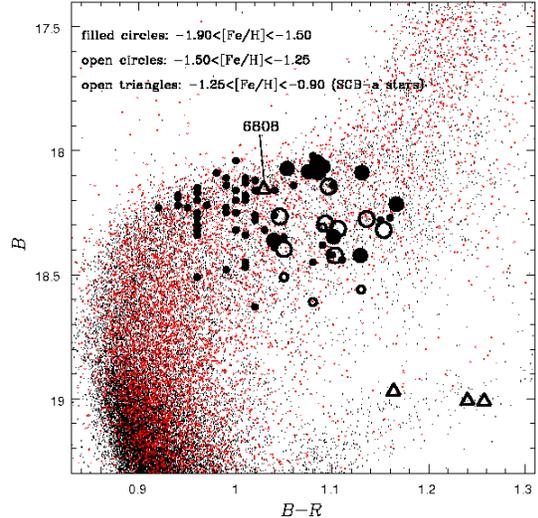}
\caption{A superposition of the CMD from the \hst\ photometry,
         transformed, according to precepts given by Sirianni et al.\
         (2005), to ground-based $B$ and $R$ (small black points), and
         the CMD from the WFI photometry (red points).  In both cases
         only a randomly selected subsample of all of the measured stars
         is plotted, in order to avoid confusion.  The symbols indicate
         the 80 stars for which we measured the [Fe/H] abundance.
         Different symbols indicate different metallicity groups, as
         indicated in the internal label.  Large symbols are \hst\
         stars, small symbols WFI stars.}
\label{f18}
\end{figure}

For each metallicity we selected from Pietrinferni et al.\ (2004, 2006)
a set of isochrones separated by 1 Gyr.  In all cases, we assumed for
the $\alpha$ enhancement the average of the [Ca/Fe] and [Ti/Fe] values
from the present study.  For stars having $-1.50 \le$ [Fe/H] $\leq
-1.25$ we assumed an enhanced helium $Y=0.38$, as was suggested in P05
for the bMS.  But for high metallicity ([Fe/H] $>-1.1$,
corresponding to the MS-a/SGB Group D/RGB-a population)
we did not assume enhanced
helium, because unlike the case of the bMS, there is no evidence for it.
Note that Sollima et al.\ (2005a) did adopt a helium enhancement for the
metal-rich population too.  In any case, because of the location of the
program stars on the flattest part of the SGB (cf.\ Figs.~\ref{f17}
and~\ref{f18}), the ages would not be strongly affected by a helium
enhancement, as can be inferred from the isochrones of Pietrinferni et
al.\ (2006); see also Fig.~6 of S05.  In fact, we have verified that for
[Fe/H] $>-1.4$ a change from $Y=0.245$ to $Y=0.35$ would move the flat
part of the isochrones by an amount that corresponds to an age
difference of only $\sim$1 Gyr.

\begin{figure}[h!]
\epsscale{1.0}
\plotone{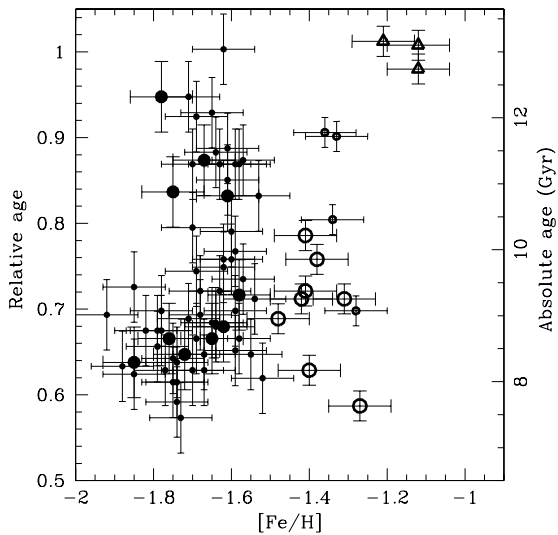}
\caption{The relative ages of the target stars, plotted against their
          [Fe/H] abundances.  The left-hand axis is fractional age
          relative to the age of the oldest stars, which was
          arbitrarily assumed to be 13 Gyr.  The right-hand axis is
          absolute age, on the same assumption.  Large symbols are
          \hst\ stars, while small ones are WFI stars.  Star 6808 is
          omitted from this figure.}
\label{f19}
\end{figure}

To estimate the age of each individual star we first shifted the
isochrones in magnitude and color by adopting the same reddening and
distance modulus that we used in Sect.~6 to derive the
atmospheric parameters of our stars.  We then calibrated the
$X$ coordinate of Fig.~\ref{f4} in terms of age, as follows.  Taking
advantage of the fact that in the SGB region all of the isochrones are
almost parallel to the the heavy line in the upper panel of
Fig.~\ref{f4},
we derived an average $X$
coordinate for each isochrone.  We then calibrated the $X$ coordinate as
a linear function of age and metallicity, which finally allowed us to
assign an age to each of our observed stars.
We do not consider these reliable as absolute ages, however, so we
converted them to relative ages, by assuming that the age
of the oldest SGB groups
corresponds to 13 Gyr (typical for a Galactic GC).  The resulting
relative ages are given in Table~\ref{t2} for the \hst\ stars, and in
Table~\ref{t3} for the WFI stars, and they are all plotted in
Fig.~\ref{f19} as a function of metallicity.

The internal errors of the relative ages shown in Fig.~\ref{f19} were
derived by adding in quadrature the effects of the photometric errors
and the uncertainties in the metallicities.
Additional errors come
from the uncertainty in He content; this may affect the ages of
intermediate-metallicity and metal-rich stars by $\sim$1--2 Gyr.

We now comment on the resulting ages as displayed in Fig.~\ref{f19}.
Both the \hst\ and the WFI samples show the same general trend.
Figure~\ref{f19} shows that the metal-poor stars have the largest age
dispersion, and split into two distinct groups, one $\sim$30\% younger
than the other.  In absolute terms, this would correspond to an age
difference of 3--4 Gyr.  Within the younger and the older metal-poor
groups the relative dispersion is consistent with what is expected for
the error budget.  Two well-defined, distinct ages for the metal-poor
stars are to be favored over an age dispersion, because this is what
the discrete distribution of the stars on the SGB would suggest (see
Figs.~\ref{f4} and \ref{f5}).  Our present spectroscopic sample is too
small to conclude firmly whether we are dealing with two really
distinct age groups, or whether there is instead a continuous
distribution in ages, but with two strong peaks at specific values.
In any case, the empirical evidence is that the metal-poor SGB stars
show age differences that correspond to a large fraction of the
cluster age.

The intermediate-metallicity SGB stars have an age dispersion that is
smaller, but still not negligible, spanning a relative age interval of
about 20\%, if we consider only the \hst\ sample.  Stars in this
sample appear to be coeval with the younger of the metal-poor SGB
groups.  Note, however, that our intermediate-metallicity SGB stars
could well include some stars with normal helium, in which case the
age spread would have been overestimated.

Finally, the most metal-rich stars, SGB Group D, turn out to be as old
as the oldest stars in the cluster, i.e., to be coeval with the oldest
metal-poor SGB stars.  Taking Fig.~\ref{f19} at face value, one can
hardly distinguish a unique age-metallicity relation.  Rather, one can
broadly distinguish four age/metallicity groups, namely, an
old metal-poor and a young metal-poor population, a
young intermediate-metallicity population, and an intriguing
old metal-rich population. The young components are some 30\% younger
than the old ones.

As a check on these relative ages estimated by isochrone fitting, we
turn to another widely used method, based on the luminosity difference
between the HB and the main sequence turn-off (see, e.g., Iben \&
Renzini 1984).  We apply this method via two different approaches: First
we use it to check the age of one of our SGB groups relative to that of
another SGB group; then we use it to compare the age of one of the
groups with that of another globular cluster.

To test the relative ages of groups, we used SGB Group A (the young
metal-poor group), and SGB Group D (the old metal-rich group); these are
the only groups for which the turn-off is clearly recognizable.  We
measured the F435W magnitudes and F435W $-$ F625W colors of their
turn-offs in our ACS photometry, and used Eq.~(12) of Sirianni et al.\
(2005), with coefficients taken from their Table 22, to derive the $V$
magnitudes of the two turn-offs.  For the HB magnitudes we used data on
RR Lyrae stars from Sollima et al.\ (2006a), who gave metallicities and
time-averaged $V$ magnitudes for a sample of 74 RR Lyrae stars in
$\omega$~Cen.  These stars make up two distinct populations, one with an
average [Fe/H] $\sim-1.7$, and the other with [Fe/H] $\sim-1.2$.  We
identify the former as the HB progeny of the SGB-MP population and the
latter as the HB progeny of our SGB Group D.  In this way we were able
to derive the magnitude difference between the HB and the TO for each of
our two selected groups.  Comparing this empirical difference with the
theoretical one derived from the models by Pietrinferni et al.\ (2004,
2006), we indeed confirm that SGB Group A appears to be $\sim 30\%$
younger than the SGB Group D population.

We also followed an alternative approach, comparing SGB Group A directly
with another GC of similar metallicity, M13.  From Sollima et al.\
(2006a) we know that the average luminosity of the metal-poor RR Lyraes
in $\omega$ Cen is $\langle V \rangle =14.4$, and we estimate 
$V=14.6\pm0.1$ for the lower envelope of the RR Lyrae distribution, which
corresponds to the zero-age HB.  In order to avoid transforming from the
ACS photometric bands of Fig.~1 to the Johnson $V$ band, we used our WFI
photometry, and estimated $V_{\rm MSTO}=18.10\pm 0.25$ for SGB Group A.
Thus we have $\Delta V^{\rm HB}_{\rm MSTO}=3.5\pm 0.3$.  This value,
though very uncertain, is typical for old globular clusters with
intermediate metallicities.  De Angeli et al.\ (2005) find a nearly
identical value of $\Delta V^{\rm HB}_{\rm MSTO}$ for M13, a cluster
whose metal content is similar to that of SGB Group A.  However, we have
argued that SGB Group A represents the youngest component in $\omega$
Cen; hence the older components would be significantly older than
typical globular clusters, possibly resulting in an age that is
embarrassingly old compared with that of the universe.  We note at the
same time that S05 needed an age of 16 Gyr to fit $\omega$ Cen (although
we are reluctant to consider absolute ages in any case).  Moreover, De
Angeli et al.\ find for the GC M3, whose metallicity is also similar, an
age only 3/4 as great as that of M13---perhaps not so surprising,
however, since M3 and M13 are a well-known ``second-parameter pair''.
This intriguing aspect requires further study, including a more accurate
estimate of $\Delta V^{\rm HB}_{\rm MSTO}$ and, most importantly, a more
thorough spectroscopic exploration of the various SGBs and HB
components.  Available spectroscopic data are still not sufficent to
explore the absolute ages of the various components of $\omega$ Cen. 

Large age differences among the stars of $\omega$~Cen are not a new
finding.  Smith et al.\ (2000), by analyzing the pattern of s-process
elements, suggested a prolonged process of star formation, lasting of
the order of 2--3 Gyr.  Hilker \& Richtler (2000, 2002) argued for an
extended star-formation period of up to 6 Gyr.
Rough estimates of ages, both from broad- and narrow-band photometry,
also show a large age spread, as presented by Hilker and Richtler
(2000), Rey et al.\ (2004), and Hughes et al.\ (2004).  An
age-metallicity relation was found by Stanford et al.\ (2003) from a
combined photometric and spectroscopic study, with a mean age difference
of 4--5 Gyr between the stars at [Fe/H] = $-$1.7 (our oldest SGB
metal-poor group) and those at [Fe/H] $\simeq -1.4$ (our youngest
intermediate-metallicity group).  Hilker et al.\ (2004) also find an age
dispersion, and they find an age-metallicity relation for stars with
[Fe/H] $<-1.0$, with a difference of about 6 Gyr between the youngest
and oldest stars in their sample;
however, the young component of the metal-poor
group is absent from their picture.

By contrast S05, using the same \hst\ images that we used here and
metallicities derived from the Ca II triplet, concluded that all groups
are substantially coeval, within the $\pm 2$-Gyr uncertainty that
affects their estimates.  Our results agree with theirs only in that
the most metal-rich component (SGB Group D = SGB-a) is found to be
coeval with the oldest metal-poor group.  Conversely, the
strongest evidence for age differences comes from the luminosity
differences among metal-poor SGB stars that we have documented here for
the first time.  We also note that S05 can argue for coeval
subpopulations only after assuming a different helium content for each of
them, rather than  only for the bMS component, as in the present study.

In closing this section, we emphasize that we refer to age
differences, as opposed to simply an age spread.  This is justified by the
clearly discrete structure of the SGB, which argues in favor of distinct
episodes of star formation rather than a continuous, albeit
fluctuating, star-formation history. We believe that this discreteness,
together with a broken age-metallicity relation (with some metal-poor
stars being younger than some metal-rich ones), puts strong
constraints on possible formation scenarios for this puzzling stellar
system.

%##########################
\section{Discussion}
%##########################

Based on deep HST photometry of the central regions of $\omega$ Cen, and
on high-resolution spectroscopy of a sample of SGB stars in the cluster,
we have identified and characterized four distinct age/metallicity
groups of stars.  These groups include over 90\% of the cluster
population in the region that we have explored; our data do not allow us
properly to characterize the remaining part.

These groups include:

\begin{enumerate}

\item
An old and metal-poor population ([Fe/H] $\sim-1.7$), with an age that
we assume to be that of the oldest GCs of the Milky Way, $\sim$13 Gyr.
This group appears to populate primarily the SGB Group C component,
and makes up about a third of the metal-poor population of the
cluster.  Note, however, that because of photometric errors it is
sometime difficult to
tell whether an individual star belongs to SGB Group B or to Group C.

\item
A young metal-poor group ([Fe/H] $\sim-1.7$), on average up to 3--4
Gyr younger than the previous group.  This group (our SGB Group A)
includes about 33\% of the $\omega$~Cen SGB stars, and makes up the
remaining two thirds of the metal-poor SGB population.

\item
A young intermediate-metallicity group, with [Fe/H] $\sim-1.4$.  It
appears to be nearly coeval with the young metal-poor
component.  This group seems to populate primarily (but not
exclusively) SGB Group B, and, because of its
metallicity, may be connected to the bMS component. If so, it should
be helium rich, and include about a third of the SGB stars in this
central region of the cluster where our observations are.

\item
An old metal-rich group for which we find [Fe/H] $\sim-1.1$, although
both its average metallicity and the extent of a possible metallicity
dispersion remain to be established better.  For the metallicity that we
have adopted, this group appears to be as old as the old metal-poor
group, i.e., with an age of the order of our assumed age for the oldest
population ($\sim$13 Gyr).  Its age would decrease by $\sim$1.5 Gyr if
we were to adopt the higher metallicity [Fe/H] = $-$0.6 that is favored
by Sollima et al.\ (2005a).  The group is confined to the region of SGB
Group D, and includes $\sim$10\% of the SGB stars in the part of
$\omega$~Cen that we have explored.  On the RGB side it connects to the
RGB-a sequence of Pancino et al.\ (2000), and on the main-sequence side
it connects to the MS-a component that we have identified in Sect.~3
(see Fig.~3).

\item
In addition, we have identified a group of stars, including about 8\% of
the cluster SGB population, that spread between SGB Groups C and D.  No
metallicity information is available for this small component, so that
we cannot say anything about its age.

\end{enumerate}

The first four groups can readily be recognized in Fig.~\ref{f19}, and
the occasional ambiguity in assigning stars to either SGB Group B or
Group C can be appreciated in Figs.~\ref{f17} and~\ref{f18}.

\subsection{Formation and evolution of the cluster}

The diversity of ages and metallicities displayed in Fig.~\ref{f19}
eliminates once and for all the possibility that $\omega$~Centauri
had its origin in a single progression of star formation
and metal enrichment.
There is no way in which, within the gravitational domain of a
globular cluster, gas masses with different chemical compositions
could have maintained separate identities
until the time of latest star formation.

The structure of Fig.~\ref{f19} bespeaks a multiplicity of origins.
Indeed, a single developing population would follow a
line that from the top left slopes downward and to the right, tracing
the increase of metallicity with time.  In Fig.~\ref{f19} only the old
metal-poor stars, joined with the younger intermediate-metallicity
stars, could satisfy this chemical-enrichment criterion, even within the
uncertainties of our ages and metallicities.  However, the most
metal-rich population, with its extreme old age, cannot fit into such a
scheme.  Rather, it must have come from a brief star-forming episode in
a different region, where enrichment had proceeded very rapidly.  By
contrast, the large clump of points at the lower left, which seems to
slope in the wrong sense, defies any simple chemical-evolution scenario.
Overall, the mix of ages and metallicities in $\omega$~Centauri suggests
multiple birth locations, followed by a later merging process.

This need for late merging seems to exclude the possibility of
explaining $\omega$~Centauri as a merger of several of the fragments
that went to make up the early Milky Way.  Our Galaxy must have been
fully formed, at least in its initial incarnation, long before the
youngest metal-poor stars of $\omega$ Cen were born.  Since the merger
of several Milky Way globular clusters seems very unlikely, because of
their small target cross-sections, we are led to the often-made
suggestion that the cluster that we now see is the remnant of a dwarf
galaxy that was captured by the Milky Way; what remains of it today
would be the central core that was bound tightly enough to resist the
tidal buffeting that removed the outer parts of the system.  Such a
central core could easily contain a broad mixture of the populations
that had made up that galaxy.  In order to allow time for the separate
development of all the populations that we see, the capture must have
taken place several billion years after the birth of the Milky Way.
The capture must have taken place far enough in the past, however, for
tidal forces and orbital spreading to have dissipated any residual
stream, such as the one in which the Sagittarius dwarf is still
embedded.
We also note that all four main age/metallicity components appear to
be enhanced in $\alpha$\ elements (see Fig.~16). Following the
canonical interpretation, this would imply that each component
underwent rapid chemical enrichment (hence remained dominated by SNII
nucleosynthetic products), in spite of the large age differences among
them.

The above conclusions seem fairly well established, but an alternative
explanation has been advanced, which should be dismissed once and for all.
It has been suggested,
e.g. by Ferraro et al.\ (2004) and by Freyhammer et al.\ (2005), that
the stars that define some of the sequences discussed above might lie at
a distance different from that of the other stars of the cluster.  It
strains credulity, however, to suggest that the mixture of populations
could be explained by a simple alignment effect, since this would
require two or more rich clusters along the same direction in the sky,
with very similar radial velocities (as pointed out by P05, and shown in
more detail in Sect.~4 of the present paper) and proper motions (whose
similarity was shown by Anderson 2003).  An alternative might be to
assume that $\omega$~Cen is a very elongated object, and that we are
looking at it along its major axis, but such an implausible structure
would not be able to survive the strong dynamical interactions that
$\omega$~Cen has with the Galaxy.  Finally, the CMD in Fig.~\ref{f1}
shows that the most metal-rich population has a MS that is clearly on
the red side of the other populations, in sharp disagreement with an
explanation of this sequence as a difference in distance.

Further insight into the formation history of $\omega$~Cen may come from
the internal distribution of the various subpopulations.  Stars with
different metal content appear to have a different radial distribution
(Norris et al.\ 1996), qualitatively consistent with the radial gradient
in the bMS/rMS ratio reported by Sollima et al.\ (2006b).  Moreover, an
asymmetric distribution with respect to the cluster center has been
suggested by Jurcsik (1998) and by Pancino et al.\ (2003).
As a separate complication,
Calamida et al.\ (2005) have claimed that $\omega$~Cen is affected by
differential reddening.  However, the sharpness of the sequences in our
Figures~\ref{f1}--\ref{f6}, especially SGB Group D and the upper edge
of the MS turn-off region, suggests that the existence of any serious
differential reddening is very unlikely.

Two more issues related to chemical abundances need to be addressed:\
the [$\alpha$/Fe] ratios of the various subpopulations, and the origin
of the helium enhancement that seems to exist in at least one of them.
All 22 stars
for which we have measured abundances of individual elements are high in
Ca and Ti, indicating [$\alpha$/Fe] $\sim$ 0.3--0.4, irrespective of
their membership in the various subpopulations
(Pancino et al.\ (2002) found [$\alpha$/Fe] $\sim$ 0.10 for
RGB-a; as we mentioned previously for [Fe/H], there is a discrepancy
between their results and ours.  We note however, that their abundance
study, like ours, included only three stars that belong to this
population.)

An alpha enhancement is generally interpreted as evidence for prompt
chemical enrichment, over a time scale shorter than the typical delay
with which Type Ia supernovae released the bulk of the iron.  This delay
is traditionally assumed to be of the order of 1 Gyr, i.e.,
significantly shorter that the age differences that we have estimated
for the various subpopulations.  The fact that all components appear to
be enhanced in $\alpha$ elements again argues for a short duration of
the enrichment processes that preceded the formation of all the cluster
subpopulations.

Following Norris (2004) and Bedin et al.\ (2004), we share the view
that the bMS component is enriched in helium.  This is
one of the most puzzling aspects of this exceptional
object.  When did the helium enrichment take place---before or after
this population merged with the rest of the body of $\omega$ Cen?  Are
other subpopulations enriched in helium?  The available data do
not allow us to venture an answer to these questions.  However, our
understanding of the formation history of this object will not be
complete as long as they remain unanswered.

Finally, we should not forget star 6808 (the brightest open triangle in
Figs.~\ref{f17} and 18).  As discussed in previous sections, we consider
its metallicity and its position in the CMD to be quite well
established.
Still, it is a quite unusual star, in that it has the
metallicity of SGB Group D but is 0.85 magnitude brighter in the $m_{\rm
F435W}$ band than any other star in this group.  It is very unlikely to
be a field-star contaminant, because it lies in the central part of the
cluster and its radial velocity and proper motion agree well with those
of the cluster.  Star 6808 could be evidence of an extreme anomaly, an
age difference of 5 or more Gyr within the metal-rich population.
Alternatively, in a less extreme interpretation we might suggest that
star 6808 is possibly a blue straggler, related to the SGB Group D
population, even though it seems {\it a priori} very unlikely for us to
have encountered one such star among only four stars that turned out to
be metal rich.  A more extended spectroscopic campaign could help to
clarify this last issue too.

\acknowledgements
The authors wish to thank the referee, John Norris,
for his careful reading of the manuscript, and for stimulating comments.
This project has been partially supported by the Italian MIUR, under the
program PRIN2003.  J.\ A.\ and A.\ M.\ C.\ acknowledge support by STScI
grant GO-9442, I.\ R.\ K.\ support by grant GO-10101.

%______________________________________________________________
%

\begin{deluxetable}{rccccccccccc}
\tabletypesize{\scriptsize}
\tablewidth{0pt}
\setlength{\tabcolsep}{0.02in}
\tablecaption{\scriptsize{Observed SGB-WFI stars.}}
\tablehead{
\colhead{ID} & \colhead{R.A.(J2000.0)} & \colhead{Decl.(J2000.0)} &
\colhead{$V$} &  \colhead{$B-V$} & \colhead{$V-I_C$} &
\colhead{$V-R_C$} & \colhead{$RV_{\rm{H}}$(km/s)} &
\colhead{$T_{\rm{eff}}(K)$} &
\colhead{$\log g$} &
\colhead{[Fe/H]} & \colhead{Age}
}
\startdata
\multicolumn{12}{c}{\small{SGB$-$MP $+$ SGB$-$MInt2 stars}}\\
12310 & 201.580363 & $-$47.653532 & 17.54 & 0.57 & 0.84 & 0.42 & 218 &
5795 & 3.9 & $-$1.75 & 0.61\\
12496 & 201.664258 & $-$47.652604 & 17.46 & 0.58 & 0.79 & 0.42 & 219 &
5807 & 3.9 & $-$1.73 & 0.57\\
14612 & 201.554592 & $-$47.641987 & 17.53 & 0.56 & 0.79 & 0.42 & 228 &
5843 & 4.0 & $-$1.74 & 0.59\\
14716 & 201.633617 & $-$47.641492 & 17.54 & 0.57 & 0.82 & 0.44 & 232 &
5775 & 3.9 & $-$1.67 & 0.63\\
15257 & 201.598488 & $-$47.638770 & 17.62 & 0.55 & 0.77 & 0.41 & 217 &
5909 & 4.0 & $-$1.77 & 0.63\\
15328 & 201.506173 & $-$47.638341 & 17.57 & 0.57 & 0.79 & 0.39 & 222 &
5896 & 4.0 & $-$1.74 & 0.61\\
16128 & 201.549683 & $-$47.634763 & 17.63 & 0.55 & 0.79 & 0.39 & 245 &
5942 & 4.0 & $-$1.52 & 0.62\\
16212 & 201.572946 & $-$47.634394 & 17.39 & 0.62 & 0.87 & 0.46 & 221 &
5594 & 3.8 & $-$1.85 & 0.62\\
16385 & 201.648017 & $-$47.633606 & 17.70 & 0.55 & 0.75 & 0.41 & 215 &
5943 & 4.1 & $-$1.65 & 0.68\\
17389 & 201.491942 & $-$47.629072 & 17.53 & 0.59 & 0.83 & 0.43 & 240 &
5759 & 3.9 & $-$1.67 & 0.65\\
17559 & 201.632454 & $-$47.628463 & 17.64 & 0.56 & 0.78 & 0.40 & 229 &
5913 & 4.0 & $-$1.59 & 0.65\\
19322 & 201.579104 & $-$47.621130 & 17.73 & 0.55 & 0.77 & 0.41 & 208 &
5906 & 4.1 & $-$1.78 & 0.70\\
19930 & 201.480925 & $-$47.618674 & 17.57 & 0.57 & 0.81 & 0.42 & 208 &
5805 & 3.9 & $-$1.88 & 0.63\\
20004 & 201.513454 & $-$47.618406 & 17.66 & 0.56 & 0.81 & 0.42 & 221 &
5839 & 4.0 & $-$1.59 & 0.70\\
20055 & 201.667229 & $-$47.618346 & 17.67 & 0.55 & 0.77 & 0.42 & 216 &
5892 & 4.0 & $-$1.78 & 0.67\\
21948 & 201.595187 & $-$47.610929 & 17.58 & 0.58 & 0.83 & 0.44 & 215 &
5748 & 3.9 & $-$1.82 & 0.67\\
23057 & 201.662250 & $-$47.606833 & 17.62 & 0.58 & 0.80 & 0.42 & 242 &
5826 & 4.0 & $-$1.64 & 0.68\\
24029 & 201.661362 & $-$47.603400 & 17.50 & 0.63 & 0.88 & 0.47 & 218 &
5559 & 3.8 & $-$1.85 & 0.73\\
24141 & 201.428288 & $-$47.602701 & 17.64 & 0.55 & 0.77 & 0.39 & 207 &
5953 & 4.0 & $-$1.70 & 0.63\\
24410 & 201.467408 & $-$47.601893 & 17.58 & 0.58 & 0.82 & 0.42 & 241 &
5780 & 3.9 & $-$1.79 & 0.66\\
24503 & 201.449554 & $-$47.601526 & 17.70 & 0.54 & 0.78 & 0.39 & 220 &
5961 & 4.1 & $-$1.74 & 0.64\\
25948 & 201.539033 & $-$47.596667 & 17.46 & 0.63 & 0.89 & 0.46 & 245 &
5575 & 3.8 & $-$1.79 & 0.67\\
6791  & 201.658221 & $-$47.687012 & 17.67 & 0.56 & 0.75 & 0.39 & 228 &
5972 & 4.1 & $-$1.58 & 0.67\\
9018  & 201.664329 & $-$47.672232 & 17.67 & 0.55 & 0.75 & 0.39 & 238 &
5993 & 4.1 & $-$1.55 & 0.65\\
10012 & 201.542692 & $-$47.665885 & 17.68 & 0.59 & 0.82 & 0.42 & 235 &
5772 & 4.0 & $-$1.62 & 0.76\\
10328 & 201.695171 & $-$47.664269 & 17.91 & 0.56 & 0.81 & 0.45 & 247 &
5787 & 4.1 & $-$1.64 & 0.88\\
11471 & 201.638833 & $-$47.658108 & 17.67 & 0.58 & 0.80 & 0.41 & 223 &
5842 & 4.0 & $-$1.54 & 0.71\\
12553 & 201.602463 & $-$47.652263 & 17.67 & 0.58 & 0.81 & 0.42 & 224 &
5811 & 4.0 & $-$1.63 & 0.72\\
15472 & 201.615946 & $-$47.637784 & 17.77 & 0.55 & 0.76 & 0.41 & 217 &
5934 & 4.1 & $-$1.57 & 0.73\\
\tablebreak
15544 & 201.709317 & $-$47.637547 & 17.67 & 0.56 & 0.77 & 0.39 & 241 &
5943 & 4.0 & $-$1.69 & 0.67\\
15948 & 201.591800 & $-$47.635583 & 17.78 & 0.56 & 0.76 & 0.40 & 226 &
5936 & 4.1 & $-$1.62 & 0.75\\
17310 & 201.606596 & $-$47.629565 & 17.67 & 0.60 & 0.79 & 0.43 & 221 &
5780 & 4.0 & $-$1.69 & 0.74\\
19573 & 201.722675 & $-$47.620179 & 17.54 & 0.60 & 0.83 & 0.41 & 241 &
5760 & 3.9 & $-$1.75 & 0.64\\
23277 & 201.765721 & $-$47.606009 & 17.68 & 0.58 & 0.87 & 0.46 & 247 &
5681 & 4.0 & $-$1.60 & 0.76\\
25395 & 201.653267 & $-$47.598729 & 17.61 & 0.59 & 0.81 & 0.42 & 218 &
5785 & 4.0 & $-$1.68 & 0.69\\
27030 & 201.709167 & $-$47.593225 & 17.74 & 0.58 & 0.82 & 0.42 & 240 &
5795 & 4.0 & $-$1.59 & 0.77\\
27176 & 201.731092 & $-$47.592774 & 17.56 & 0.59 & 0.86 & 0.44 & 216 &
5694 & 3.9 & $-$1.71 & 0.69\\
6308  & 201.663075 & $-$47.690455 & 17.73 & 0.57 & 0.73 & 0.39 & 220 &
5967 & 4.1 & $-$1.68 & 0.72\\
7037  & 201.578883 & $-$47.685244 & 17.58 & 0.60 & 0.79 & 0.41 & 228 &
5812 & 4.0 & $-$1.64 & 0.68\\
10822 & 201.571267 & $-$47.661535 & 17.94 & 0.57 & 0.76 & 0.39 & 232 &
5927 & 4.1 & $-$1.70 & 0.87\\
12382 & 201.719504 & $-$47.653239 & 18.03 & 0.60 & 0.80 & 0.42 & 239 &
5788 & 4.1 & $-$1.62 & 1.00\\
13612 & 201.661137 & $-$47.646816 & 17.92 & 0.65 & 0.86 & 0.49 & 244 &
5569 & 4.0 & $-$1.36 & 0.85\\
13906 & 201.720854 & $-$47.645374 & 17.71 & 0.62 & 0.82 & 0.42 & 231 &
5740 & 4.0 & $-$1.70 & 0.80\\
17912 & 201.651671 & $-$47.627033 & 17.62 & 0.65 & 0.90 & 0.50 & 228 &
5504 & 3.9 & $-$1.63 & 0.87\\
18508 & 201.648763 & $-$47.624502 & 17.89 & 0.62 & 0.81 & 0.42 & 229 &
5770 & 4.1 & $-$1.34 & 0.77\\
19626 & 201.576500 & $-$47.619906 & 17.80 & 0.60 & 0.83 & 0.45 & 238 &
5714 & 4.0 & $-$1.61 & 0.85\\
19778 & 201.626646 & $-$47.619421 & 17.75 & 0.63 & 0.85 & 0.46 & 242 &
5620 & 4.0 & $-$1.61 & 0.89\\
20268 & 201.638696 & $-$47.617553 & 17.59 & 0.68 & 0.90 & 0.48 & 220 &
5463 & 3.8 & $-$1.59 & 0.87\\
23171 & 201.728758 & $-$47.606451 & 17.53 & 0.61 & 0.89 & 0.45 & 240 &
5619 & 3.9 & $-$1.92 & 0.69\\
24735 & 201.579258 & $-$47.600892 & 17.83 & 0.63 & 0.83 & 0.46 & 230 &
5641 & 4.0 & $-$1.65 & 0.93\\
25115 & 201.674296 & $-$47.599680 & 17.83 & 0.60 & 0.88 & 0.50 & 242 &
5567 & 4.0 & $-$1.71 & 0.95\\
25250 & 201.487533 & $-$47.598979 & 17.77 & 0.65 & 0.87 & 0.45 & 231 &
5572 & 4.0 & $-$1.69 & 0.92\\
28869 & 201.584737 & $-$47.587474 & 17.85 & 0.60 & 0.80 & 0.41 & 235 &
5808 & 4.1 & $-$1.58 & 0.87\\
6493  & 201.671692 & $-$47.689141 & 17.89 & 0.58 & 0.77 & 0.41 & 230 &
5873 & 4.1 & $-$1.57 & 0.87\\
8307  & 201.662962 & $-$47.676813 & 17.73 & 0.61 & 0.80 & 0.40 & 242 &
5820 & 4.0 & $-$1.60 & 0.79\\
8675  & 201.539787 & $-$47.674428 & 17.95 & 0.66 & 0.82 & 0.42 & 235 &
5694 & 4.1 & $-$1.33 & 0.85\\
9271  & 201.597104 & $-$47.670620 & 17.74 & 0.61 & 0.83 & 0.44 & 225 &
5719 & 4.0 & $-$1.53 & 0.83\\
9352  & 201.569362 & $-$47.670094 & 17.66 & 0.65 & 0.82 & 0.44 & 222 &
5694 & 4.0 & $-$1.28 & 0.68\\
\enddata
\label{t3}
\end{deluxetable}

\begin{deluxetable}{ccc}
\tabletypesize{\scriptsize}
\tablewidth{0pt}
\tablecaption{\scriptsize{Identified stellar populations}}
\tablehead{
\colhead{MS} & \colhead{SGB} & \colhead{RGB}
}
\startdata
\hline
rMS  & SGB Group A & RGB-MP \\
57\% & 33\%  & 42$\pm$8\% \\
$-$1.7 & $-$1.7  & $-$1.7 \\[5pt]
bMS  & SGB Group B & RGB-MInt1 \\
33\% & 29\%  & 28$\pm$6\% \\
$-$1.4 & $-$1.7 and $-$1.4 & $-$1.5 \\[5pt]
\    & SGB Group C & RGB-MInt2 \\
\    & 20\%  & 17$\pm$5\% \\
\    & $-$1.7 and $-$1.4 & $-$1.2 \\[5pt]
\    &  ?    & RGB-MInt3 \\
\    &  8\%  & 8$\pm$3\% \\
\    & $-$1.1? & $-$1.0 \\[5pt]
MS-a & SGB Group D & RGB-a \\
10\%(?) & 10\% & 5$\pm$1\% \\
$-$1.1 and/or $-$0.6 & $-$1.1 and/or $-$0.6 & $-$0.8 \\
\enddata
\label{t4}
\end{deluxetable}

\end{document}